\documentclass{article}

\usepackage[preprint, nonatbib]{neurips_2024}

\usepackage[utf8]{inputenc}             
\usepackage[T1]{fontenc}                
\usepackage{hyperref}                   
\usepackage{url}                        
\usepackage{booktabs}                   
\usepackage{multirow}                   
\usepackage{amsfonts}                   
\usepackage{nicefrac}                   
\usepackage{microtype}                  
\usepackage{enumitem}                   
\usepackage{amsmath}                    
\usepackage{graphicx}                   
\usepackage{wrapfig}                    
\usepackage[normalem]{ulem}             
\useunder{\uline}{\ul}{}                
\usepackage[table,xcdraw]{xcolor} 
\usepackage{arydshln}                   
\usepackage{caption}                    

\title{Mercury: A Code Efficiency Benchmark for Code Large Language Models}

\author{
    Mingzhe Du$^{1,2}$, Anh Tuan Luu$^{1}$, Bin Ji$^{2}$, Qian Liu$^{3}$, See-Kiong Ng$^{2}$ \\
    \textsuperscript{1}Nanyang Technological University\\
    \textsuperscript{2}National University of Singapore\\
    \textsuperscript{3}Sea AI Lab\\
    {\texttt \{mingzhe001, anhtuan.luu\}@ntu.edu.sg, \{jibin, seekiong\}@nus.edu.sg, liuqian@sea.com}
}

\begin{document}

\maketitle

\begin{abstract}
    Amidst the recent strides in evaluating Large Language Models for Code~(Code LLMs), existing benchmarks have mainly focused on the functional correctness of generated code, neglecting the importance of their computational efficiency. To fill the gap, we present \emph{Mercury}, the first code efficiency benchmark for Code LLMs. It comprises 1,889 Python tasks, each accompanied by adequate solutions that serve as real-world efficiency baselines, enabling a comprehensive analysis of the runtime distribution.
    Based on the distribution, we introduce a new metric \texttt{Beyond}, which computes a runtime-percentile-weighted \texttt{Pass} score to reflect functional correctness and code efficiency simultaneously. On \emph{Mercury}, leading Code LLMs can achieve 65\% on \texttt{Pass}, while less than 50\% on \texttt{Beyond}. Given that an ideal \texttt{Beyond} score would be aligned with the \texttt{Pass} score, it indicates that while Code LLMs exhibit impressive capabilities in generating functionally correct code, there remains a notable gap in their efficiency. Finally, our empirical experiments reveal that Direct Preference Optimization~(DPO) serves as a robust baseline for enhancing code efficiency compared with Supervised Fine Tuning~(SFT), which paves a promising avenue for future exploration of efficient code generation.~\footnote{Our code and data are available on GitHub: \href{https://github.com/Elfsong/Mercury}{https://github.com/Elfsong/Mercury}.}
\end{abstract}

\section{Introduction}
The domain of code generation, which aims to empower computers to autonomously generate code based on natural language task descriptions~(NL2Code), has long been considered a promising way to facilitate interaction between humans and computers~\cite{zan2022large, wong2023natural}. 
The recent emergence of Large Language Models~(LLMs) has spurred a new wave of NL2Code models~\cite{openai2023gpt4,roziere2023code,deepseekcoder,lozhkov2024starcoder,qwen}, which leverage the impressive language understanding and generative capabilities of LLMs to drive forward the ambitious goal of synthesizing high-quality code from natural language instructions.

To measure the quality of code, recent code generation benchmarks mainly focus on evaluating their functional correctness via test case fuzzing~\cite{liu2024your}. This approach assesses the outcome congruence between the LLM-generated and canonical solutions by executing bespoken test cases. For instance, HumanEval~\cite{chen2021evaluating} and MBPP~\cite{austin2021program} collected a small but fine set of handcrafted tasks with test cases. EvalPlus~\cite{evalplusliu} further consolidates these two above benchmarks by augmenting the case scope. On the contrary, APPS~\cite{hendrycks2021measuring} widely gathered over 5,000 public coding tasks from online platforms. 
Despite these strides, there is a discernible oversight in current code generation benchmarks concerning the code efficiency evaluation, although that is critical in software development~\cite{xu2022survey, zan2023large}. Moreover, handcrafting diverse solutions and test cases to cover all scenarios is infeasible~\cite{evalplusliu}. In light of these findings, we highlight vital limitations inherent in the existing code generation benchmarks:

\begin{figure}[ht]
    \centering
    \includegraphics[width=0.96\textwidth]{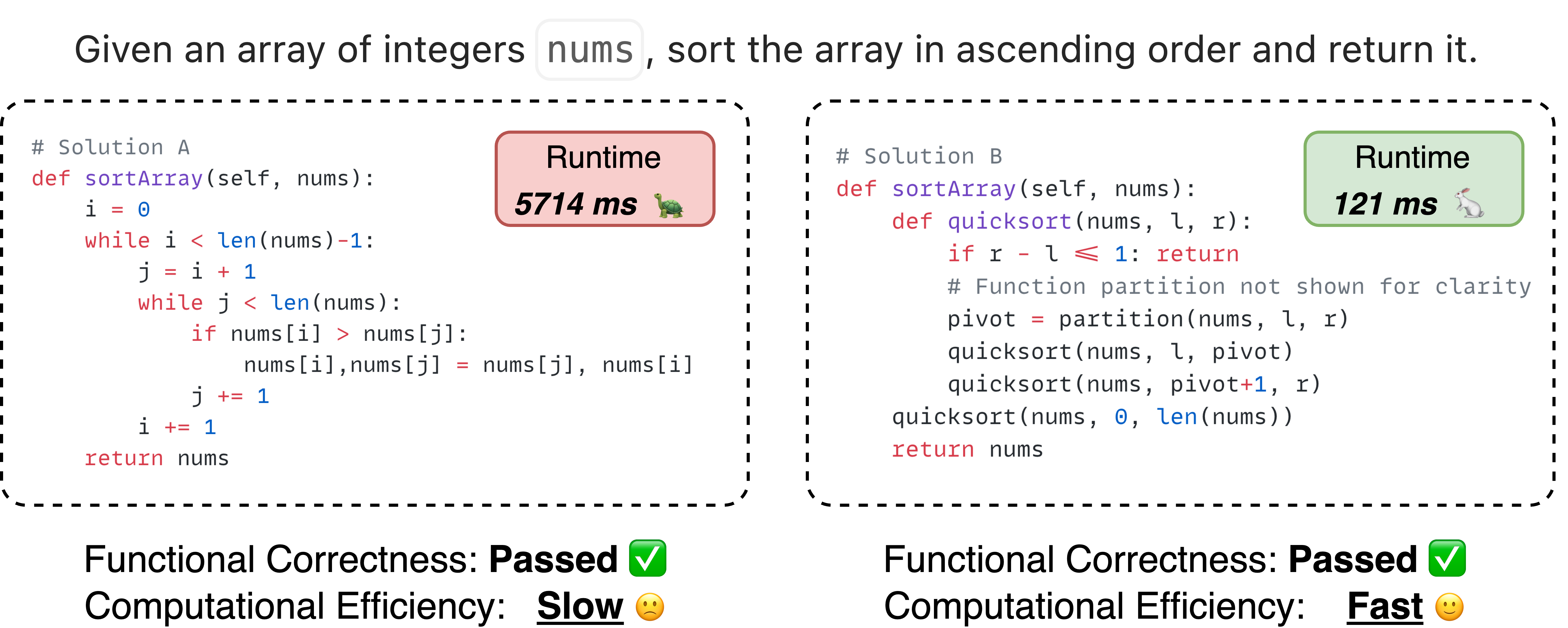}
    \caption{
        Executing these two LLM-generated codes on 100 test cases.
        While both codes successfully follow the task instruction and pass all test cases, the \emph{right} snippet notably excels in code efficiency, completing in a mere 121~ms compared to the 5,714~ms consumed by the \emph{left} snippet. As Code LLMs become widely used in the real world, code efficiency determines factual productivity, where Mercury can gauge the vital metric.
    }
    \label{fig:mercury_motivation}
\end{figure}

\begin{enumerate}[leftmargin=*]
    \item \textbf{Absence of code Efficiency Evaluation.} 
    Existing code generation benchmarks focus on assessing functional correctness while overlooking the evaluation of code efficiency~\cite{chen2021evaluating, austin2021program, hendrycks2021measuring}. As illustrated in Figure~\ref{fig:mercury_motivation}, despite both code snippets can handle the sorting task functionally, the \emph{right} efficient solution~(\textit{121~ms}) is nearly 50 times faster than the \emph{left} inefficient solution~(\textit{5,714~ms}). 
    This striking runtime differentiation underscores the necessity of incorporating code efficiency assessments within code generation benchmarks, encouraging Code LLMs to produce not only correct but also efficient code.

    \item \textbf{Insufficient Test Case Coverage.} 
    As shown in Table~\ref{tab:data_comparsion}, most code generation benchmarks manually build a small number of test cases or extract the accompanying test cases from existing resources, potentially overlooking edge cases and nuanced code behaviors~\cite{chen2021evaluating, austin2021program}. For example, Figure~\ref{fig:mercury_comparison} displays that \textit{HumanEval \#55} contains only 3 test cases, testing up to the 12th Fibonacci number~\cite{chen2021evaluating}. Its given canonical solution will quickly reach the recursion depth limitation when computing a larger Fibonacci number (the recursion limitation depends on the environment). Therefore, notwithstanding the generated code satisfies all test cases, such success does not necessarily equate to assurance of functional correctness and much less to code efficiency.

    \item \textbf{Lack of Task Diversity.} 
    Another noticeable deficit of existing code generation benchmarks is the insufficient diversity and complexity in their tasks~\cite{chen2021evaluating, austin2021program, liu2024your}. Since most benchmarks only consist of elementary-level programming tasks, recent Code LLMs can effortlessly tackle most tasks regardless of their actual capacities~\cite{zan2023large}. This flaw results in these benchmarks failing to pose a substantial challenge to Code LLMs and truly reflect their underlying potential. 
\end{enumerate}

\textbf{Code Efficiency.} 
\emph{Code efficiency refers to the performance measure of time and space complexity to accomplish a specific task.} 
Efficient code can improve user experience, save energy, and make applications more sustainable and cost-effective. Compared with the scalable memory space, execution time is the performance bottleneck of most codes. Consequently, this work focuses on the time dimension of code efficiency. 

\begin{figure*}[t]
    \centering
    \includegraphics[width=\textwidth]{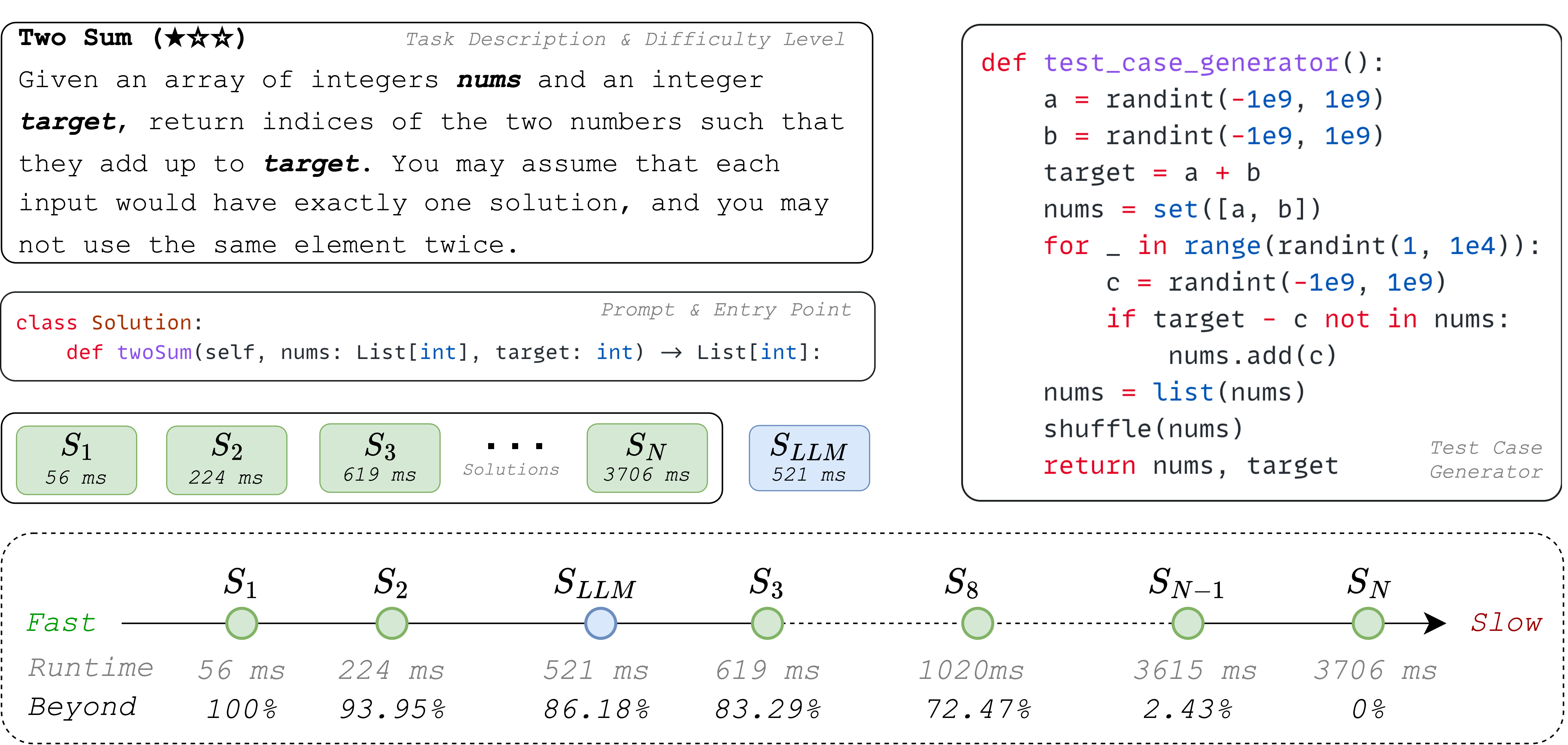}
    \caption{
        An overview of \emph{Mercury} dataset. Each \emph{Mercury} task has a task description, a test case generator, a prompt \& entry point, and corresponding solutions. To evaluate code efficiency, we introduce the Beyond metric, which signifies the runtime percentile of the LLM-generated code on the runtime distribution supported by corresponding solutions. 
        In this example, the LLM-generated code executes in 521 ms, outpacing 86.18\% of collected solutions on the runtime distribution. Consequently, the \texttt{Beyond} metric in this case is 86.18\%.    }
    \label{fig:mercury_overview}
\end{figure*}

\textbf{Our Benchmark.} In this work, we introduce \emph{Mercury}, a novel code generation benchmark designed to assess and improve the code efficiency of Code LLMs. As depicted in Figure~\ref{fig:mercury_overview}, \emph{Mercury} comprises 1,889 Python programming tasks with three difficulty stratification, which is divided into two datasets for model evaluation and fine-tuning separately. For each evaluation task, we assign a test case generator to remedy the shortfall of test case coverage. In measuring code efficiency, the primary challenge stems from normalizing the absolute runtime across tasks that have diverse runtime ranges. Thus, we collect and locally execute numerous historical solutions for each task to form a runtime distribution and leverage the runtime percentile of LLM-generated code on the distribution instead of the absolute runtime to evaluate code efficiency. Furthermore, to mitigate performance discrepancies attributed to irrelevant processes and diverse hardware configurations, we set up an isolated sandbox environment for task execution to establish local runtime distributions.
    
\textbf{Contribution.} Our work aimed to fill the code efficiency evaluation gap in code generation benchmarks with the following key contributions:
\begin{itemize}[leftmargin=*]
    \item \textbf{Dataset.} We collect a novel code generation dataset \emph{Mercury} designed to assess and improve Code LLM code efficiency in Section~\ref{sec:datasets}, accompanied by an extensible open-source data collection framework for enriching \emph{Mercury} with more tasks and programming languages. 
    \item \textbf{Metric.} We propose the first efficiency-focused code generation metric \texttt{Beyond} and establish a benchmark to evaluate leading Code LLMs using this metric in Section~\ref{sec:metric}.
    \item \textbf{Baselines.} In Section~\ref{sec:experiments}, we detail our extensive analysis of two baselines to enhance code efficiency while maintaining functional correctness. Experiment results reveal that despite Code LLMs excelling in functional correctness, there is still considerable potential to elevate efficiency.
\end{itemize}

\begin{table}[htbp]
    \caption{
        A comparison of \emph{Mercury} to existing NL2Code benchmarks. \emph{Mercury} distinguishes itself by including a set of distilled high-quality solutions and a dedicated test case generator for each task. * signifies that the solution number can be further expanded by the data collection framework.
        \vspace{0.2em}
    }
    \label{tab:data_comparsion}
    \centering
    \begin{tabular}{lccccc}
        \toprule
        \textbf{Benchmarks}     & \textbf{Tasks} & \textbf{Sources} & \textbf{Cases}   & \textbf{Solutions}     & \textbf{Difficulty} \\ \midrule
        HumanEval               & 164            & Crowd Source     & 8.08             & 1                      & 1                   \\
        MBPP                    & 257            & Crowd Source     & 3.01             & 1                      & 1                   \\
        APPS                    & 5,000          & Online           & 21.2             & 23.4                   & 3                   \\
        \textbf{Mercury}        & 256            & Online + Filters & +$\infty$        & 18.4 *                 & 3                   \\ \bottomrule
    \end{tabular}
\end{table}

\section{Mercury Datasets}
\label{sec:datasets}
We initiate this work by collecting public programming tasks on Leetcode~\cite{leetcode}. Subjecting these questions to a series of filters, we distilled them down to 1,889 high-quality tasks. A difficulty-balanced subset of 256 tasks was randomly selected to form the \textbf{Mercury-eval} benchmark, which obtains an average of 18.4 solutions for each problem. The remaining tasks have been designated as the \textbf{Mercury-train} dataset for baseline training (detailed data distribution is listed in Appendix Table~\ref{tab:dataset_distribution}). To enhance clarity within this paper, we employ \emph{Mercury} to denote \textbf{Mercury-eval} unless otherwise specified.

\textbf{Data Schema.} 
As illustrated in Figure~\ref{fig:mercury_overview}, \emph{Mercury} offers a unified data schema to streamline the evaluation procedure and bolster further development endeavors. The data scheme encompasses these principal components: 
(1) \textbf{Task Description} contains the task instruction interpreted into a plain text format, along with illustrative examples and constraints of inputs and outputs.
(2) \textbf{Test Case Generator} refers to a Python code snippet designed to automatically produce a comprehensive set of test cases in accordance with the specifications laid out in the task description. 
(3) \textbf{Solutions} are sampled from Leetcode historical submissions. Each solution within Mercury has undergone rigorous testing, and Locality-Sensitive Hashing~\cite{jafari2021survey} is employed to prevent the inclusion of any identical solutions.
(4) \textbf{Prompts and Entry Points} where prompts act as the initiating prefixes for LLM code generation and entry points denote the start point for code execution.
We delineate the definition of \emph{Mercury} fields in the Appendix Table~\ref{tab:mercury_fields}.

\textbf{Task Filters.}
\emph{Mercury} tasks originate from public programming problems on Leetcode. To assure the quality and uniformity of the dataset, we distilled gathered tasks based on the following conditions: 

\begin{enumerate}[leftmargin=*]
    \vspace{-0.3em}
    \item \textbf{Number of Solutions.} 
    To establish a solution runtime distribution for each task, we filtered out tasks having less than two associated solutions. After excluding these tasks, \emph{Mercury} tasks possess an average of 18.4 unique solutions.
    \vspace{-0.3em}
    \item \textbf{Restricted Data Structure.} 
    Above the inherent Python data types, \emph{Mercury} also incorporates two custom data types: Binary Tree and Linked List (the specific structure definitions can be found in Appendix Figure~\ref{fig:mercury_data_structure}), which increases \emph{Mercury's} diversity and escalates its difficulty level. Tasks that contain other data structures will be removed.
    \vspace{-0.3em}
    \item \textbf{Unique Outputs.} Certain Leetcode tasks may permit non-unique answers. For example, a result list can be returned in any order. Evaluating all possible answers can drastically complicate the test case verification process. To eliminate this problem, we harness the corresponding test case generator to generate $N$ test cases $T_i = \langle Input_i, Output_i\rangle \ s.t. \ i\in\{0, 1, \cdots, N\}$ and execute $T$ on different solutions $S_m \ s.t. \ m\in\{0, 1, \cdots, M\}$ to observe if all $Output_i = S_m(Input_i) \ s.t. \ i\in\{0, 1, \cdots, N\}$ remain identical. Any tasks that potentially yield non-unique answers were subsequently excluded. 
\end{enumerate}

\textbf{Task Difficulty.} 
Most existing NL2Code benchmarks predominantly comprise simplistic tasks, leading to a situation where LLMs of varied capabilities address most tasks effortlessly and yield indistinguishable high scores~\cite{zan2023large, jain2024livecodebench}. To alleviate this issue, \emph{Mercury} inherits the difficulty categorization from Leetcode, \textit{i.e.}, Easy, Medium, and Hard. The stratification aims to probe the upper bounds of Code LLM capabilities, delivering a more evident distinction between various Code LLMs.

\textbf{Test Case Generator.} 
Manual creation of test cases can be a laborious process. To gather sufficient test cases to conduct an exhaustive assessment, we assign a test case generator for each evaluation task, which can produce a full range of test cases to thoroughly evaluate the functional correctness and code efficiency of given solutions.
Specifically, We feed $pretty\_content$ into GPT-4~\cite{openai2023gpt4} to generate an initial test case generator snippet. To confirm the effectiveness of the initial generator, we subsequently create 24 test cases by the generator and submit these cases to the Leetcode Online Judge (OJ) system. Should any of the generated test cases not pass the LeetCode OJ validation, we manually revise the generator until all generated cases can be successfully validated.

\section{Code Efficiency Metric}
\label{sec:metric}
In the domain of software development, code efficiency can be defined as the absolute code runtime for executing a given test case set~\cite{chen2022learning}. Nonetheless, a primary obstacle in benchmarking code efficiency is normalizing runtime measurements across disparate environments. For instance, a sub-optimal solution might have a faster absolute runtime on high-performance hardware than an optimal solution on low-performance hardware. 
Moreover, different operation systems and code interpreters may also fluctuate the code runtime. Therefore, absolute runtime fails as a consistent and reliable code efficiency benchmark metric. To address this issue, an intuitive approach involves modeling a devoted runtime distribution for each task and calculating the average runtime percentiles of LLM solution samples over the runtime distribution. With this idea in mind, we proposed a normalized code efficiency metric \texttt{Beyond}:

\begin{equation}
    p_{k}^{n} = \frac{max(R^{n}) - clip(r_{k}^{n}, min(R^{n}), max(R^{n}))}{max(R^{n}) - min(R^{n})},
    \qquad
    Beyond = \frac{\sum\nolimits_{N, K}^{n=0, k=0} p_{k}^{n} }{N \cdot K}.
    \label{eq:beyond_metric}
\end{equation}


\noindent Where $N$ is the total number of tasks, and $K$ denotes the size of LLM solution samples. For a specific task $n \in N$, $R^{n}$ is the runtime array corresponding to the collected historical solutions, and $r_k^{n} \ s.t. \ k \in K$ denotes the runtime for the $k$-th LLM solution. 
$clip$ is a function to constraint the value $r_{k}^{n}$ in the range $[min(R^{n}), max(R^{n})]$.
\emph{Runtime} is defined as the period from the solution instantiation to the evaluation across all test cases, culminating with a successful termination (More engineering details can be found in Appendix Section~\ref{sec:sandbox_details}). 
Since any case failure of the $k$-th solution results in $r_{k}^{n} \rightarrow +\infty$ and then \emph{$p_{k}^{n} = 0$}, \texttt{Beyond} can reflect functional correctness as well.

\noindent \textbf{Untrusted Code Execution.}
Since most Code LLMs are trained on an extensive code corpus from unverified sources, there is an intrinsic risk that these models may produce malicious code when driven by specific meticulous prompts~\cite{chen2021evaluating}. The direct execution of synthesized code raises significant security concerns. To alleviate the risk of running untrusted code, we engage a robust sandbox to execute code in an isolated environment. Sandbox details are deliberated in Appendix~\ref{sec:sandbox_details}

\noindent \textbf{Environment-agnostic Evaluation.} To ensure fair comparison across diverse configurations, we run each task $n$ with corresponding test cases locally and aggregate their runtimes into the runtime array $R^{n}$. Appendix Figure~\ref{fig:hardware_independent} illustrates the \texttt{Beyond} score of two LLMs (`deepseek-coder-33b' and `deepseek-coder-6.7b') over three distinct hardware specifications: the micro-tier (0.25 CPU cores), the small-tier (0.5 CPU cores), and the standard-tier (1 CPU core). The results demonstrate that \emph{Beyond} remains consistent over different hardware configurations.

\vspace{-0.6em}
\section{Experiments}
\vspace{-0.6em}
\label{sec:experiments}
In this section, we present a series of baseline experiments to improve code efficiency by training on \emph{Mercury-train} dataset and assessing on the \emph{Mercury-eval} dataset. Our empirical study encompasses 10 open-source LLMs with a broad parameter spectrum from 1.3 to 34 billion. For each LLM, we compare the performance of the original model and two optimization strategies, Supervised Fine-Tuning~(SFT) and Direct Preference Optimization~(DPO), for their potential to optimize LLM generating functionally correct and computationally efficient code. Finally, we analyzed the underlying factors contributing to the failure of LLMs on the \emph{Mercury-eval} dataset.

\vspace{-0.4em}
\subsection{Baselines}
\vspace{-0.4em}
\noindent \textbf{Supervised Fine-Tunning.}
Within the SFT~\cite{bakker2022fine} method, an LLM undergoes additional training on a small dataset, which aims to specialize the LLM to perform better on certain tasks correlated to the training dataset. To optimize the code efficiency performance of Code LLMs, the most intuitive strategy is to fine-tune the Code LLM using optimal runtime solutions. In our experimental setup, we apply a unified prompt template for each Code LLM to ensure a fair comparison. The ``pretty\_content'' attribute fills the \textbf{<task\_content>} placeholder, the ``prompt'' attribute fills the \textbf{<code\_starter>} placeholder, and the \textbf{<code\_completion>} placeholder is completed with the fastest solutions. To steer Code LLMs towards generating the intended code completion format, we prepend a one-shot example to the prompt template. Appendix Figure~\ref{fig:model_prompt} presents the prompt template.

\noindent \textbf{Direct Preference Optimization.}
Although SFT exemplifies a straightforward approach, it is susceptible to the pitfall of catastrophic forgetting~\cite{kir2017overcome}. To enable LLMs to align with human preferences while preserving their functional capabilities, existing methodologies employ reinforcement learning with human preference feedback~(RLHF). However, RLHF introduces additional model complexities and potential instabilities, necessitating significant computing resources and extra reward model training~\cite{ziegler2019fine, bai2022training, stiennon2020learning}. DPO~\cite{rafailov2023direct} bypasses these challenges by explicitly mapping reward functions and the optimal objective. 
This connection demonstrates that maximizing rewards under specific constraints can be effectively addressed through a singular training phase based on data reflecting human preferences. 
The DPO training procedure is elaborated in Appendix Section~\ref{sec:dpo_details}.

\vspace{-0.8em}
\subsection{Functional Correctness Benchmarks}
\vspace{-0.4em}
\textbf{HumanEval} assesses the functional correctness of synthesized code derived from docstrings. It contains 164 distinct Python tasks that cover several programming areas, such as language comprehension, algorithm development, and simple mathematics~\cite{chen2021evaluating}. \textbf{MBPP} has a sanitized collection of 257 entry-level Python programming problems. Each problem in this dataset consists of three components: a task description, an associated code solution, and three automated test cases to validate the code functionality~\cite{austin2021program}. Both HumanEval and MBPP harness the metric \texttt{Pass} to measure the Code LLMs' functional correctness, where a task is considered solved if the given solution passes all test cases, and the total fraction of solved tasks is reported as $Pass = N_{solved} / N_{total}$~\cite{kulal2019spoc}.

\subsection{Experimental Setups}
\label{sec:experimental_setups}

\noindent \textbf{Configuration.} We employ LoRA~\cite{hu2021lora} for both SFT and DPO experiments. We set $lora\_alpha = 16$, $lora\_dropout=0.05$, and $lora\_r=8$. The optimizer is \emph{Adamw}~\cite{loshchilov2017decoupled}, and the learning rate is 1$e$-4 and 5$e$-5 for SFT and DPO, respectively. For SFT experiments, we train each model in 200 steps. For DPO experiments, we set $\beta = 0.1$ and $training\_step = 500$. For code generation, we set the temperature as 0.2. For the $Beyond$ metric calculation, we set $K = 5$. All experiments are conducted on two A100-80G GPUs. We employed {Accelerate}~\cite{accelerate} for distributed training, {DeepSpeed}~\cite{aminabadi2022deepspeed} for gradient partitioning, and {BitsandBytes}~\cite{dettmers2022llmint8} for model quantization.

\noindent \textbf{Training Data.} We use \emph{Mercury-train} for model training. As for the SFT process, we nominate the fastest solution as the supervised label, then format the training data as  $\langle pretty\_content, prompt, solution\_optimal\rangle$. Regarding the DPO procedure, we select the top 5 pairs of solutions that exhibit the most significant discrepancy in runtime. The training date format is $\langle pretty\_content, prompt, solution\_fast, solution\_slow \rangle$. 

\vspace{-0.5em}
\subsection{Empirical Results}
\vspace{-0.5em}
Functional correctness is the prerequisite for evaluating code efficiency for the code generation task. Our primary objective is to enhance the code efficiency without compromising the functional correctness. 
To this end, we first introduce the existing metric \texttt{Pass} to gauge the functional correctness~\cite{chen2021evaluating} and then leverage \texttt{Beyond} to provide a holistic evaluation, encompassing both code efficiency and functional correctness. 
Finally, we measure the \texttt{Gap} between \texttt{Beyond} and \texttt{Pass} to mirror the baseline ability for improving efficiency while preserving correctness. 
These experiments aim to investigate the innate capabilities of cutting-edge Code LLMs and their potential after baseline fine-tuning. Therefore, extensive parameter optimization and prompt engineering were not pursued. To deliver a comprehensive evaluation, we have further integrated the HumanEval and MBPP benchmarks as supplementary measures for appraising functional correctness~\cite{chen2021evaluating, austin2021program}.

\begin{figure}[h]
    \centering
    \begin{minipage}{0.48\linewidth}
        \includegraphics[width=\linewidth]{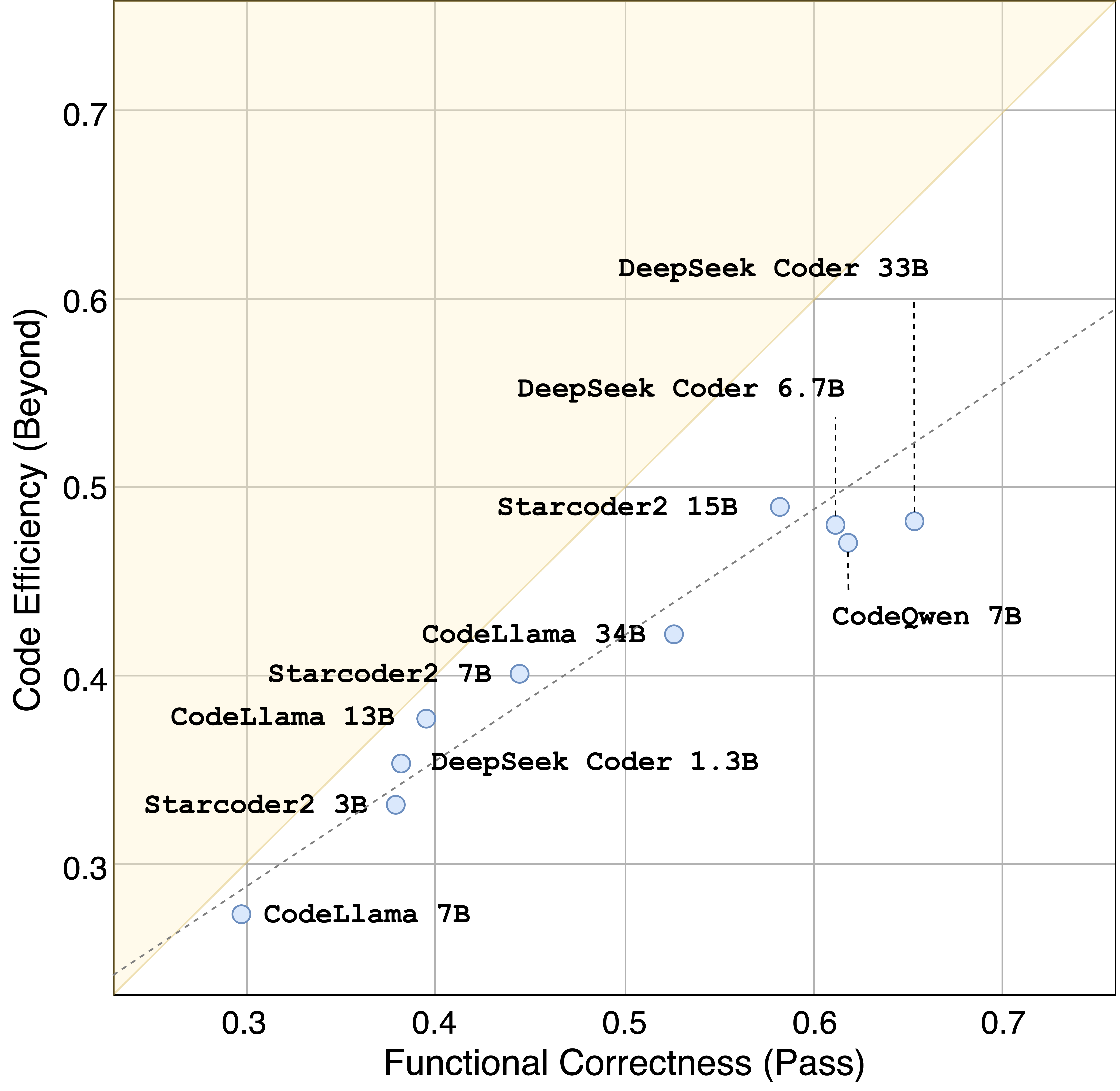} 
    \end{minipage}
    \hfill
    \begin{minipage}{0.48\linewidth}
        \includegraphics[width=\linewidth]{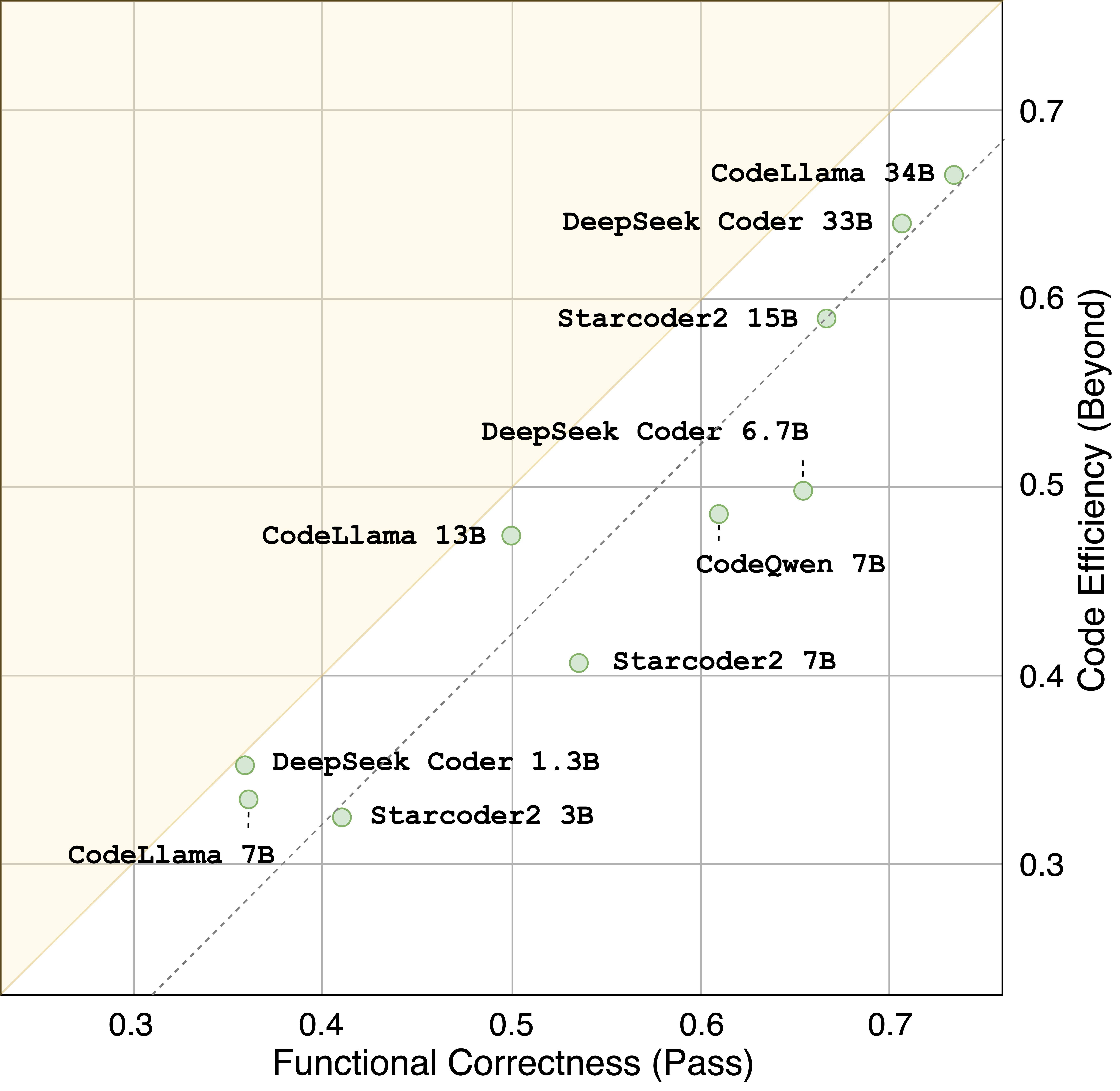}
    \end{minipage}
    \caption{
        The horizontal axis represents the score for functional correctness, while the vertical axis indicates the score for code efficiency.
        The \emph{left} figure illustrates the performance of the baseline model, whereas the \emph{right} one depicts the performance after DPO tuning.
        Model points located nearer to the diagonal line exhibit a more equitable balance between functionality and efficiency.
        \vspace{-1em}
    }
    \label{fig:visualization}
\end{figure}

\noindent \textbf{Functional Correctness.}
Table~\ref{tab:functional_correctness_metrics} lists \texttt{Pass}~scores over various Code LLMs, showing that larger models tend to provide better functional correctness. Except for the smallest model ``deepseek-coder-1.3b-base'', DPO invariably enhances the overall \texttt{Pass} scores across most Code LLMs, while SFT diminishes functional correctness on the largest two Code LLMs. 
These findings suggest that smaller models may struggle to integrate new knowledge while preserving their original functionality, and SFT may induce catastrophic forgetting in the pursuit of heightened code efficiency. Moreover, it is evident on \emph{Mercury} that \texttt{Pass} scores of each model consistently decline as the difficulty level increases, indicating that the \emph{Mercury} difficulty stratification is effective at probing the upper limitation of each Code LLM compared to the auxiliary benchmarks.

\begin{table}[ht]
    \centering
    \tabcolsep=10pt
    \caption{
        Functional correctness~(\texttt{Pass}) evaluation results. \underline{The underlined values} denote the top-performed approaches among the original model and baselines. \textbf{The bolded values} denote the best performance on each benchmark. We sample one solution for each task to calculate \emph{pass} score. 
    }
    \label{tab:functional_correctness_metrics}
    \renewcommand{\arraystretch}{0.7}
    \resizebox{\textwidth}{!}{
        \begin{tabular}{lcccccl}
            \toprule
            \multicolumn{1}{c}{}                                               &                                      &                                 & \multicolumn{3}{c}{\textbf{Mercury}}              &                                                                                               \\ \cline{4-6}
            \multicolumn{1}{c}{\multirow{-2}{*}{\textbf{Model Name}}}          & \multirow{-2}{*}{\textbf{HumanEval}} & \multirow{-2}{*}{\textbf{MBPP}} & \textbf{Easy}             & \textbf{Medium}       & \textbf{Hard}         & \multirow{-2}{*}{\textbf{Overall}}                                    \\ \midrule
            \textbf{deepseek-coder-1.3b-base}                                  & 28.7                                 & {\underline{55.4}}                      & 60.7                      & 52.8                  & 23.2                  & \cellcolor{blue!10} {\underline{38.1}}                                        \\
            + SFT                                                              & 24.2                                 & 46.2                            & 58.9                      & {\underline{53.6}}            & {\underline{25.3}}            & \cellcolor{red!10} 36.2 {\color[HTML]{680100}(-1.9)}                  \\
            + DPO                                                              & {\underline{29.1}}                           & 50.2                            & {\underline{61.4}}                & {\underline{53.6}}            & 20.0                  & \cellcolor{red!10} 35.9 {\color[HTML]{680100}(-2.2)}                  \\ \midrule
            \textbf{starcoder2-3b}                                             & 31.7                                 & 57.4                            & 56.1                      & 52.1                  & 21.6                  & \cellcolor{blue!10} 37.8                                              \\
            + SFT                                                              & 29.0                                 & 47.2                            & 60.7                      & 58.8                  & {\underline{25.3}}            & \cellcolor{green!10} 38.8 {\color[HTML]{036400}(+1.0)}                \\
            + DPO                                                              & {\underline{33.5}}                           & {\underline{59.6}}                      & {\underline{62.5}}                & {\underline{61.0}}            & 23.4                  & \cellcolor{green!10} {\underline{41.1}} {\color[HTML]{036400}(+3.3)}          \\ \midrule
            \textbf{deepseek-coder-6.7b-base}                                  & 47.6                                 & 70.2                            & 69.3                      & 68.9                  & 56.1                  & \cellcolor{blue!10} 61.0                                              \\
            + SFT                                                              & {\underline{56.1}}                           & 59.6                            & 69.1                      & 71.4                  & 57.7                  & \cellcolor{green!10} 62.2 {\color[HTML]{036400}(+1.2)}                \\
            + DPO                                                              & 54.3                                 & {\underline{72.8}}                      & {\underline{74.1}}                & {\underline{72.6}}            & {\underline{58.9}}            & \cellcolor{green!10} {\underline{65.4}} {\color[HTML]{036400} (+4.4)}         \\ \midrule
            \textbf{starcoder2-7b}                                             & 35.2                                 & 54.4                            & 63.6                      & 61.7                  & 29.2                  & \cellcolor{blue!10} 44.3                                              \\
            + SFT                                                              & 42.9                                 & 57.2                            & 64.8                      & 58.5                  & 31.3                  & \cellcolor{green!10} 47.5 {\color[HTML]{036400}(+3.2)}                \\
            + DPO                                                              & 55.4                                 & 61.4                            & 74.8                      & 66.9                  & 32.6                  & \cellcolor{green!10} {\underline{53.6}} {\color[HTML]{036400}(+9.3)}          \\ \midrule
            \textbf{CodeLlama-7b-hf}                                           & 33.5                                 & {\underline{52.0}}                      & 55.7                      & 41.7                  & 12.9                  & \cellcolor{blue!10} 29.6                                              \\
            + SFT                                                              & 29.5                                 & 47.6                            & 58.9                      & 38.5                  & 16.1                  & \cellcolor{green!10} 31.3 {\color[HTML]{036400}(+1.7)}                \\
            + DPO                                                              & {\underline{38.7}}                           & 49.2                            & {\underline{67.5}}                & {\underline{45.7}}            & {\underline{17.9}}            & \cellcolor{green!10} {\underline{36.1}} {\color[HTML]{036400}(+6.5)}          \\ \midrule
            \textbf{CodeQwen1.5-7B}                                            & 51.8                                 & 72.2                            & 70.0                      & {\underline{70.1}}            & {\underline{49.7}}            & \cellcolor{blue!10} 61.1                                              \\
            + SFT                                                              & 54.3                                 & 74.8                            & 70.9                      & 67.9                  & {\underline{49.7}}            & \cellcolor{green!10} {\underline{61.9}} {\color[HTML]{036400}(+0.8)}          \\
            + DPO                                                              & {\underline{55.5}}                           & {\underline{75.4}}                      & {\underline{72.5}}                & 66.9                  & 45.7                  & \cellcolor{green!10} 61.1 {\color[HTML]{036400}(+0)}                  \\ \midrule
            \textbf{starcoder2-15b}                                            & 46.3                                 & 66.2                            & 69.5                      & 65.4                  & 50.3                  & \cellcolor{blue!10} 58.0                                              \\
            + SFT                                                              & 51.6                                 & 69.2                            & 72.0                      & 68.9                  & 51.7                  & \cellcolor{green!10} 61.3 {\color[HTML]{036400}(+3.3)}                \\
            + DPO                                                              & {\underline{57.0}}                           & {\underline{72.8}}                      & {\underline{78.0}}                & {\underline{73.8}}            & {\underline{54.7}}            & \cellcolor{green!10} {\underline{66.7}} {\color[HTML]{036400}(+8.7}          \\ \midrule
            \textbf{CodeLlama-13b-hf}                                          & 37.8                                 & 62.4                            & 76.8                      & {\underline{60.5}}            & 18.4                  & \cellcolor{blue!10} 39.6                                              \\
            + SFT                                                              & 39.5                                 & 59.8                            & 65.5                      & 54.8                  & 19.5                  & \cellcolor{red!10} 39.5 {\color[HTML]{680100} (-0.1)}                 \\
            + DPO                                                              & {\underline{49.1}}                           & {\underline{64.4}}                      & {\underline{78.6}}                & 60.0                  & {\underline{29.0}}            & \cellcolor{green!10} {\underline{50.1}} {\color[HTML]{036400} (+10.6)}        \\ \midrule
            \textbf{deepseek-coder-33b-base}                                   & 54.3                                 & 73.2                            & 70.9                      & 67.9                  & 62.3                  & \cellcolor{blue!10} 65.0                                              \\
            + SFT                                                              & 58.1                                 & 74.8                            & 61.8                      & 58.0                  & 47.1                  & \cellcolor{red!10} 58.7 {\color[HTML]{680100}(-6.3)}                  \\
            + DPO                                                              & \textbf{{\underline{72.9}}}                  & \textbf{{\underline{80.6}}}             & {\underline{78.9}}                & \textbf{{\underline{76.5}}}   & {\underline{61.6}}            & \cellcolor{green!10} \textbf{{\underline{73.4}}} {\color[HTML]{036400}(+8.4)} \\ \midrule
            \textbf{CodeLlama-34b-hf}                                          & 48.2                                 & 65.4                            & 77.7                      & 63.7                  & 32.4                  & \cellcolor{blue!10} 52.4                                              \\
            + SFT                                                              & 52.8                                 & 68.2                            & 61.8                      & 58.0                  & 26.2                  & \cellcolor{red!10} 47.5 {\color[HTML]{680100}(-4.9)}                  \\
            + DPO                                                              & {\underline{65.9}}                           & {\underline{75.2}}                      & \textbf{{\underline{83.9}}}       & {\underline{68.4}}            & \textbf{{\underline{63.2}}}   & \cellcolor{green!10} {\underline{70.6}} {\color[HTML]{036400}(+18.2)}         \\ \bottomrule
        \end{tabular}
    }\vspace{-1.5em}
\end{table}

\begin{table}[ht!]
    \centering
    \tabcolsep=10pt
    \caption{
        Code efficiency (\texttt{Beyond}) evaluation results across three difficulty levels. \textbf{The bolded value} indicates the top performance for each metric, while \underline{the underlined values} denote the most effective approaches among the original model and the baselines.
        In our experiment, we sample 5 solutions for each task to calculate \emph{Beyond} score.
    }
    \label{tab:code_efficiency_metrics}
    \renewcommand{\arraystretch}{0.82}
    
    \resizebox{\textwidth}{!}{
        \begin{tabular}{lcccll}
            \toprule
            \textbf{Model name}               & \textbf{Easy}               & \textbf{Medium}           & \textbf{Hard}             & \textbf{Overall}                                                          & \textbf{Gap}                                                              \\ \midrule
            \textbf{deepseek-coder-1.3b-base} & {\underline{47.97}}                 & 39.77                     & {\underline{19.26}}               & \cellcolor{blue!10} {\underline{35.62}}                                           & \cellcolor{blue!10} 9.85                                                  \\
            + SFT                             & 42.58                       & 38.12                     & 18.67                     & \cellcolor{red!10} 33.04 {\color[HTML]{680100}(-2.58)}                    & \cellcolor{red!8} 12.74 {\color[HTML]{680100}(+2.89)}                     \\
            + DPO                             & 46.91                       & {\underline{42.27}}               & 16.78                     & \cellcolor{red!10} 35.21 {\color[HTML]{680100}(-0.41)}                    & \cellcolor{green!8} {\underline{9.64}} {\color[HTML]{036400}(-0.21)}              \\ \midrule
            \textbf{starcoder2-3b}            & 43.55                       & 41.91                     & 15.21                     & \cellcolor{blue!10} 33.40                                                 & \cellcolor{blue!8} {\underline{9.72}}                                             \\
            + SFT                             & {\underline{44.64}}                 & {\underline{42.10}}               & {\underline{15.72}}               & \cellcolor{green!10} {\underline{34.01}} {\color[HTML]{036400}(+0.61)}            & \cellcolor{red!8} 14.04 {\color[HTML]{680100}(+4.31)}                      \\
            + DPO                             & 43.70                       & 41.02                     & 12.99                     & \cellcolor{red!10} 32.42 {\color[HTML]{680100}(-0.99)}                    & \cellcolor{red!8} 16.33 {\color[HTML]{680100}(+6.61)}                      \\ \midrule
            \textbf{deepseek-coder-6.7b-base} & 48.80                       & 51.16                     & {\underline{45.11}}               & \cellcolor {blue!10}48.29                                                 & \cellcolor{blue!8} {\underline{16.40}}                                            \\
            + SFT                             & 51.37                       & 52.71                     & 44.28                     & \cellcolor{green!10} 49.39 {\color[HTML]{036400}(+1.09)}                  & \cellcolor{red!8} 16.55 {\color[HTML]{680100}(+0.16)}                      \\
            + DPO                             & {\underline{56.25}}                 & {\underline{52.35}}               & 40.62                     & \cellcolor{green!10} {\underline{49.70}} {\color[HTML]{036400}(+1.41)}            & \cellcolor{red!8} 18.73 {\color[HTML]{680100}(+2.34)}                      \\ \midrule
            \textbf{starcoder2-7b}            & 50.23                       & 51.29                     & 20.25                     & \cellcolor{blue!10} 40.37                                                 & \cellcolor{blue!8} {\underline{10.95}}                                            \\
            + SFT                             & 42.21                       & 44.02                     & {\underline{21.09}}               & \cellcolor{red!10} 35.61 {\color[HTML]{680100}(-4.77)}                    & \cellcolor{red!8} 15.80 {\color[HTML]{680100}(+4.84)}                      \\
            + DPO                             & {\underline{53.52}}                 & {\underline{51.41}}               & 17.35                     & \cellcolor{green!10} {\underline{40.56}} {\color[HTML]{036400}(+0.18)}            & \cellcolor{red!8} 17.41 {\color[HTML]{680100}(+6.46)}                      \\ \midrule
            \textbf{CodeLlama-7b-hf}          & 42.55                       & 30.99                     & 8.88                      & \cellcolor{blue!10} 27.45                                                 & \cellcolor{blue!8} {\underline{9.27}}                                             \\
            + SFT                             & 39.75                       & 26.89                     & 9.55                      & \cellcolor{red!10} 25.41 {\color[HTML]{680100}(-2.04)}                    & \cellcolor{red!8} 12.48 {\color[HTML]{680100}(+3.21)}                      \\
            + DPO                             & {\underline{54.14}}                 & {\underline{34.48}}               & {\underline{11.10}}               & \cellcolor{green!10} {\underline{33.29}} {\color[HTML]{036400}(+5.84)}            & \cellcolor{green!8} 10.46 {\color[HTML]{680100}(1.19)}                    \\ \midrule
            \textbf{CodeQwen1.5-7B}           & 51.11                       & {\underline{53.56}}               & {\underline{39.03}}               & \cellcolor{blue!10} 47.78                                                 & \cellcolor{blue!8} 15.35                                                  \\
            + SFT                             & 54.16                       & 51.43                     & 38.05                     & \cellcolor{green!10} 47.82 {\color[HTML]{036400}(0.04)}                   & \cellcolor{green!8} 14.91 {\color[HTML]{036400}(-0.43)}                   \\
            + DPO                             & {\underline{56.07}}                 & 51.55                     & 38.05                     & \cellcolor{green!10} {\underline{48.52}} {\color[HTML]{036400}(0.74)}             & \cellcolor{green!8} {\underline{13.12}} {\color[HTML]{036400}(-2.22)}             \\ \midrule
            \textbf{starcoder2-15b}           & 58.18                       & 52.09                     & 37.34                     & \cellcolor{blue!10} 49.17                                                 & \cellcolor{blue!8} 12.55                                                  \\
            + SFT                             & 53.54                       & 52.77                     & 37.73                     & \cellcolor{red!10} 47.92 {\color[HTML]{680100}(-1.25)}                    & \cellcolor{red!8} 16.22 {\color[HTML]{680100}(+3.67)}                      \\
            + DPO                             & {\underline{68.29}}                 & {\underline{59.54}}               & {\underline{48.97}}               & \cellcolor{green!10} {\underline{58.95}} {\color[HTML]{036400}(9.78)}             & \cellcolor{green!8} {\underline{10.81}} {\color[HTML]{036400}(-1.74)}             \\ \midrule
            \textbf{CodeLlama-13b-hf}         & 57.00                       & 44.25                     & 12.99                     & \cellcolor{blue!10} 38.01                                                 & \cellcolor{blue!8} 13.79                                                  \\
            + SFT                             & 44.95                       & 39.96                     & 13.55                     & \cellcolor{red!10} 32.70 {\color[HTML]{680100}(-5.31)}                    & \cellcolor{green!8} 13.78 {\color[HTML]{036400}(-0.01)}                   \\
            + DPO                             & {\underline{67.09}}                 & {\underline{55.72}}               & {\underline{19.72}}               & \cellcolor{green!10} {\underline{47.39}} {\color[HTML]{036400}(9.38)}             & \cellcolor{green!8} {\underline{8.47}} {\color[HTML]{036400}(-5.32)}              \\ \midrule
            \textbf{deepseek-coder-33b-base}  & 51.26                       & 48.90                     & 45.43                     & \cellcolor{blue!10} 48.53                                                 & \cellcolor{blue!8} 18.50                                                  \\
            + SFT                             & 40.33                       & 37.75                     & 36.82                     & \cellcolor{red!10} 38.32 {\color[HTML]{680100}(-10.21)}                   & \cellcolor{green!8} 17.30 {\color[HTML]{036400}(-1.20)}                   \\
            + DPO                             & {\underline{74.59}}                 & {\underline{68.91}}               & \textbf{{\underline{55.98}}}      & \cellcolor{green!10} \textbf{{\underline{66.47}}} {\color[HTML]{036400}(+17.94)}  & \cellcolor{green!8} \textbf{{\underline{5.79}}} {\color[HTML]{036400}(-12.70)}    \\ \midrule
            \textbf{CodeLlama-34b-hf}         & 56.28                       & 48.21                     & 22.96                     & \cellcolor{blue!10} 42.40                                                 & \cellcolor{blue!8} 15.49                                                  \\
            + SFT                             & 45.49                       & 44.96                     & 20.73                     & \cellcolor{red!10} 36.91 {\color[HTML]{680100}(-5.50)}                    & \cellcolor{green!8} 11.61 {\color[HTML]{036400}(-3.88)}                   \\
            + DPO                             & \textbf{{\underline{78.55}}}        & \textbf{{\underline{60.95}}}      & {\underline{51.94}}               & \cellcolor{green!10} {\underline{63.94}} {\color[HTML]{036400}(+21.54)}           & \cellcolor{green!8} {\underline{8.01}} {\color[HTML]{036400}(-7.47)}              \\ \bottomrule
        \end{tabular}
    }\vspace{-1em}
\end{table}

\noindent \textbf{Code Efficiency.}
Regarding the NL2Code task, once functional correctness has been assured, attention naturally pivots to enhancing code efficiency. As depicted in Table~\ref{tab:code_efficiency_metrics}, we investigate code efficiency metrics across a spectrum of Code LLMs. 
Experiments demonstrate that DPO yields a stable enhancement in code efficiency from models exceeding 6.7B parameters. In contrast, SFT detracts most \texttt{Beyond} scores from original models, suggesting that the plain SFT may not be a feasible strategy for enhancing code efficiency. Further analysis compares the \texttt{Gap} between \texttt{Beyond} and $Pass$. Since the ideal \texttt{Beyond} should be aligned with \texttt{Pass} (where the LLM-generated solution is correct and faster than all historical solutions), it shows how much the baseline method shrinks the gap between functional correctness and code efficiency. Our findings indicate that DPO substantially narrows \texttt{Gap} in models larger than 15B parameters. However, \texttt{Gap} tends to widen in smaller models under the same configuration. This implies that larger models possess a greater capacity to assimilate the nuanced knowledge to make strides in code efficiency while retaining their functional correctness.

\vspace{-0.5em}
\subsection{Failure Analysis}
\vspace{-0.5em}
Table~\ref{tab:failure_analysis} provides an error breakdown of where Code LLMs misstep during the \emph{Mercury} evaluation: 

(1) \textbf{Generation Errors} arise from syntactical issues. The common manifestations include \emph{improper indentation}, \emph{mismatched parentheses}, or \emph{unexpected truncation}. Fine-tuning introduces additional knowledge for Code LLMs to adapt the Mercury convention, emphasizing standard indentation, concise code, and minimal comments. Therefore, both SFT and DPO generally reduced these errors.

(2) \textbf{Execution Errors} differ from Generation Errors because they occur after the code has been successfully loaded. These errors emerge as exceptions, which could stem from various issues, such as flawed code logic, execution timeouts, memory leakage, or sandbox interruption. We observe that SFT tends to aggravate these errors on most models, whereas DPO mitigates these errors successfully.

(3) \textbf{Test Case Errors} are the most prevalent errors where the code is executed without exceptions, but the output fails to align with the expectation. 

\begin{table}[ht]
    \centering
    \caption{
        The distribution of failure cases across \emph{Code Generation}, \emph{Code Execution}, and \emph{Test Case} errors. E/M/H indicates Easy/Medium/Hard levels, respectively. We sample 5 solutions for each task, so there are $256 * 5 = 1280$ solutions in total for each model.
    }
    \label{tab:failure_analysis}
    \renewcommand{\arraystretch}{0.92}
    \resizebox{\textwidth}{!}{
        \begin{tabular}{lcccccccccccc}
            \toprule
            \multicolumn{1}{c}{\multirow{2}{*}{\textbf{Model Name}}} & \multicolumn{3}{c}{\textbf{Code Generation}} & \multicolumn{3}{c}{\textbf{Code Execution}} & \multicolumn{3}{c}{\textbf{Test Case}} & \multicolumn{3}{c}{\textbf{Pass}} \\
            \cmidrule(lr){2-4} \cmidrule(lr){5-7} \cmidrule(lr){8-10} \cmidrule(lr){11-13} 
            \multicolumn{1}{c}{}                                     & E                                 & M                                 & H                                & E                               & M                               & H                                & E                               & M                               & H                              & E                            & M                            & H                            \\ \midrule
            \textbf{deepseek-coder-1.3b-base}                        & \cellcolor{blue!10} 82            & \cellcolor{blue!10} 85            & \cellcolor{blue!10} 59           & \cellcolor{blue!6} 17           & \cellcolor{blue!6} 33           & \cellcolor{blue!6} 95            & \cellcolor{blue!10} 74          & \cellcolor{blue!10} 73          & \cellcolor{blue!10} 180        & \cellcolor{blue!6} 267       & \cellcolor{blue!6} 214       & \cellcolor{blue!6} 101       \\
            + SFT                                                    & \cellcolor{red!10} 106            & \cellcolor{red!10} 104            & \cellcolor{green!10} 37          & \cellcolor{green!6} 16          & \cellcolor{green!6} 8           & \cellcolor{green!6} 63           & \cellcolor{green!10} 59         & \cellcolor{red!10} 76           & \cellcolor{red!10} 225         & \cellcolor{red!6} 259        & \cellcolor{green!6} 217      & \cellcolor{green!6} 110      \\
            + DPO                                                    & \cellcolor{red!10} 90             & \cellcolor{red!10} 108            & \cellcolor{green!10} 40          & \cellcolor{green!6} 17          & \cellcolor{green!6} 5           & \cellcolor{green!6} 49           & \cellcolor{green!10} 63         & \cellcolor{red!10} 75           & \cellcolor{red!10} 259         & \cellcolor{green!6} 270      & \cellcolor{green!6} 217      & \cellcolor{red!6} 87         \\ \midrule
            \textbf{starcoder2-3b}                                   & \cellcolor{blue!10} 107           & \cellcolor{blue!10} 102           & \cellcolor{blue!10} 33           & \cellcolor{blue!6} 35           & \cellcolor{blue!6} 26           & \cellcolor{blue!6} 107           & \cellcolor{blue!10} 51          & \cellcolor{blue!10} 66          & \cellcolor{blue!10} 201        & \cellcolor{blue!6} 247       & \cellcolor{blue!6} 211       & \cellcolor{blue!6} 94        \\
            + SFT                                                    & \cellcolor{green!10} 97           & \cellcolor{green!10} 93           & \cellcolor{green!10} 24          & \cellcolor{green!6} 29          & \cellcolor{green!6} 13          & \cellcolor{green!6} 90           & \cellcolor{green!10} 47         & \cellcolor{green!10} 61         & \cellcolor{red!10} 211         & \cellcolor{green!6} 267      & \cellcolor{green!6} 238      & \cellcolor{green!6} 110      \\
            + DPO                                                    & \cellcolor{green!10} 79           & \cellcolor{green!10} 75           & \cellcolor{green!10} 11          & \cellcolor{green!6} 30          & \cellcolor{green!6} 14          & \cellcolor{green!6} 87           & \cellcolor{green!10} 56         & \cellcolor{red!10} 69           & \cellcolor{red!10} 235         & \cellcolor{green!6} 275      & \cellcolor{green!6} 247      & \cellcolor{green!6} 102      \\ \midrule
            \textbf{deepseek-coder-6.7b-base}                        & \cellcolor{blue!10} 107           & \cellcolor{blue!10} 101           & \cellcolor{blue!10} 30           & \cellcolor{blue!6} 16           & \cellcolor{blue!6} 5            & \cellcolor{blue!6} 56            & \cellcolor{blue!10} 12          & \cellcolor{blue!10} 20          & \cellcolor{blue!10} 105        & \cellcolor{blue!6} 305       & \cellcolor{blue!6} 279       & \cellcolor{blue!6} 244       \\
            + SFT                                                    & \cellcolor{green!10} 105          & \cellcolor{green!10} 100          & \cellcolor{green!10} 25          & \cellcolor{red!6} 17            & \cellcolor{red!6} 6             & \cellcolor{red!6} 61             & \cellcolor{red!10} 14           & \cellcolor{green!10} 10         & \cellcolor{green!10} 98        & \cellcolor{red!6} 304        & \cellcolor{green!6} 289      & \cellcolor{green!6} 251      \\
            + DPO                                                    & \cellcolor{green!10} 87           & \cellcolor{green!10} 82           & \cellcolor{green!10} 23          & \cellcolor{green!6} 12          & \cellcolor{red!6} 6             & \cellcolor{red!6} 58             & \cellcolor{red!10} 15           & \cellcolor{red!10} 23           & \cellcolor{green!10} 98        & \cellcolor{green!6} 326      & \cellcolor{green!6} 294      & \cellcolor{green!6} 256      \\ \midrule
            \textbf{starcoder2-7b}                                   & \cellcolor{blue!10} 107           & \cellcolor{blue!10} 101           & \cellcolor{blue!10} 22           & \cellcolor{blue!6} 21           & \cellcolor{blue!6} 9            & \cellcolor{blue!6} 74            & \cellcolor{blue!10} 32          & \cellcolor{blue!10} 45          & \cellcolor{blue!10} 212        & \cellcolor{blue!6} 280       & \cellcolor{blue!6} 250       & \cellcolor{blue!6} 127       \\
            + SFT                                                    & \cellcolor{green!10} 105          & \cellcolor{green!10} 100          & \cellcolor{green!10} 22          & \cellcolor{green!6} 18          & \cellcolor{red!6} 13            & \cellcolor{green!6} 72           & \cellcolor{green!10} 32         & \cellcolor{red!10} 55           & \cellcolor{green!10} 205       & \cellcolor{green!6} 285      & \cellcolor{red!6} 237        & \cellcolor{green!6} 136      \\
            + DPO                                                    & \cellcolor{green!10} 90           & \cellcolor{green!10} 90           & \cellcolor{green!10} 21          & \cellcolor{green!6} 10          & \cellcolor{red!6} 11            & \cellcolor{green!6} 61           & \cellcolor{green!10} 11         & \cellcolor{green!10} 33         & \cellcolor{green!10} 211       & \cellcolor{green!6} 329      & \cellcolor{green!6} 271      & \cellcolor{green!6} 142      \\ \midrule
            \textbf{CodeLlama-7b-hf}                                 & \cellcolor{blue!10} 23            & \cellcolor{blue!10} 28            & \cellcolor{blue!10} 23           & \cellcolor{blue!6} 41           & \cellcolor{blue!6} 69           & \cellcolor{blue!6} 122           & \cellcolor{blue!10} 131         & \cellcolor{blue!10} 139         & \cellcolor{blue!10} 234        & \cellcolor{blue!6} 245       & \cellcolor{blue!6} 169       & \cellcolor{blue!6} 56        \\
            + SFT                                                    & \cellcolor{green!10} 11           & \cellcolor{green!10} 9            & \cellcolor{green!10} 17          & \cellcolor{red!6} 44            & \cellcolor{red!6} 72            & \cellcolor{green!6} 112          & \cellcolor{green!10} 126        & \cellcolor{red!10} 168          & \cellcolor{red!10} 236         & \cellcolor{green!6} 259      & \cellcolor{red!6} 156        & \cellcolor{green!6} 70       \\
            + DPO                                                    & \cellcolor{green!10} 9            & \cellcolor{green!10} 10           & \cellcolor{green!10} 12          & \cellcolor{green!6} 23          & \cellcolor{green!6} 56          & \cellcolor{green!6} 117          & \cellcolor{green!10} 111        & \cellcolor{red!10} 154          & \cellcolor{green!10} 228       & \cellcolor{green!6} 297      & \cellcolor{green!6} 185      & \cellcolor{green!6} 78       \\ \midrule
            \textbf{CodeQwen1.5-7B}                                  & \cellcolor{blue!10} 105           & \cellcolor{blue!10} 100           & \cellcolor{blue!10} 18           & \cellcolor{blue!6} 1            & \cellcolor{blue!6} 4            & \cellcolor{blue!6} 44            & \cellcolor{blue!10} 26          & \cellcolor{blue!10} 17          & \cellcolor{blue!10} 157        & \cellcolor{blue!6} 308       & \cellcolor{blue!6} 284       & \cellcolor{blue!6} 216       \\
            + SFT                                                    & \cellcolor{green!10} 105          & \cellcolor{red!10} 101            & \cellcolor{red!10} 16            & \cellcolor{red!6} 4             & \cellcolor{green!6} 3           & \cellcolor{green!6} 35           & \cellcolor{green!10} 19         & \cellcolor{red!10} 26           & \cellcolor{red!10} 168         & \cellcolor{green!6} 312      & \cellcolor{red!6} 275        & \cellcolor{green!6} 216      \\
            + DPO                                                    & \cellcolor{green!10} 98           & \cellcolor{green!10} 96           & \cellcolor{red!10} 26            & \cellcolor{red!6} 5             & \cellcolor{red!6} 8             & \cellcolor{green!6} 35           & \cellcolor{green!10} 18         & \cellcolor{red!10} 30           & \cellcolor{red!10} 175         & \cellcolor{green!6} 319      & \cellcolor{red!6} 271        & \cellcolor{red!6} 199        \\ \midrule
            \textbf{starcoder2-15b}                                  & \cellcolor{blue!10} 105           & \cellcolor{blue!10} 100           & \cellcolor{blue!10} 20           & \cellcolor{blue!6} 4            & \cellcolor{blue!6} 7            & \cellcolor{blue!6} 49            & \cellcolor{blue!10} 25          & \cellcolor{blue!10} 33          & \cellcolor{blue!10} 147        & \cellcolor{blue!6} 306       & \cellcolor{blue!6} 265       & \cellcolor{blue!6} 219       \\
            + SFT                                                    & \cellcolor{green!10} 104          & \cellcolor{green!10} 100          & \cellcolor{green!10} 18          & \cellcolor{green!6} 3           & \cellcolor{green!6} 3           & \cellcolor{red!6} 56             & \cellcolor{green!10} 16         & \cellcolor{green!10} 23         & \cellcolor{green!10} 136       & \cellcolor{green!6} 317      & \cellcolor{green!6} 279      & \cellcolor{green!6} 225      \\
            + DPO                                                    & \cellcolor{green!10} 83           & \cellcolor{green!10} 64           & \cellcolor{green!10} 10          & \cellcolor{green!6} 1           & \cellcolor{green!6} 1           & \cellcolor{green!6} 33           & \cellcolor{green!10} 13         & \cellcolor{red!10} 41           & \cellcolor{green!10} 141       & \cellcolor{green!6} 343      & \cellcolor{green!6} 299      & \cellcolor{green!6} 251      \\ \midrule
            \textbf{CodeLlama-13b-hf}                                & \cellcolor{blue!10} 10            & \cellcolor{blue!10} 14            & \cellcolor{blue!10} 28           & \cellcolor{blue!6} 19           & \cellcolor{blue!6} 41           & \cellcolor{blue!6} 99            & \cellcolor{blue!10} 73          & \cellcolor{blue!10} 105         & \cellcolor{blue!10} 228        & \cellcolor{blue!6} 338       & \cellcolor{blue!6} 245       & \cellcolor{blue!6} 80        \\
            + SFT                                                    & \cellcolor{red!10} 46             & \cellcolor{blue!10} 52            & \cellcolor{red!10} 32            & \cellcolor{red!6} 29            & \cellcolor{green!6} 19          & \cellcolor{red!6} 111            & \cellcolor{red!10} 77           & \cellcolor{red!10} 112          & \cellcolor{green!10} 207       & \cellcolor{red!6} 288        & \cellcolor{red!6} 222        & \cellcolor{green!6} 85       \\
            + DPO                                                    & \cellcolor{red!10} 24             & \cellcolor{green!10} 9            & \cellcolor{green!10} 24          & \cellcolor{green!6} 10          & \cellcolor{green!6} 12          & \cellcolor{red!6} 100            & \cellcolor{green!10} 60         & \cellcolor{red!10} 141          & \cellcolor{green!10} 185       & \cellcolor{green!6} 346      & \cellcolor{red!6} 243        & \cellcolor{green!6} 126      \\ \midrule
            \textbf{deepseek-coder-33b-base}                         & \cellcolor{blue!10} 105           & \cellcolor{blue!10} 103           & \cellcolor{blue!10} 26           & \cellcolor{blue!6} 11           & \cellcolor{blue!6} 11           & \cellcolor{blue!6} 47            & \cellcolor{blue!10} 12          & \cellcolor{blue!10} 16          & \cellcolor{blue!10} 91         & \cellcolor{blue!6} 312       & \cellcolor{blue!6} 275       & \cellcolor{blue!6} 271       \\
            + SFT                                                    & \cellcolor{green!10} 69           & \cellcolor{green!10} 78           & \cellcolor{red!10} 27            & \cellcolor{red!6} 27            & \cellcolor{red!6} 26            & \cellcolor{red!6} 65             & \cellcolor{red!10} 72           & \cellcolor{red!10} 66           & \cellcolor{red!10} 138         & \cellcolor{red!6} 272        & \cellcolor{red!6} 235        & \cellcolor{red!6} 205        \\
            + DPO                                                    & \cellcolor{green!10} 56           & \cellcolor{green!10} 75           & \cellcolor{green!10} 15          & \cellcolor{green!6} 9           & \cellcolor{green!6} 7           & \cellcolor{red!6} 86             & \cellcolor{red!10} 28           & \cellcolor{green!10} 13         & \cellcolor{green!10} 66        & \cellcolor{green!6} 347      & \cellcolor{green!6} 310      & \cellcolor{red!6} 268        \\ \midrule
            \textbf{CodeLlama-34b-hf}                                & \cellcolor{blue!10} 22            & \cellcolor{blue!10} 35            & \cellcolor{blue!10} 50           & \cellcolor{blue!6} 28           & \cellcolor{blue!6} 55           & \cellcolor{blue!6} 84            & \cellcolor{blue!10} 48          & \cellcolor{blue!10} 57          & \cellcolor{blue!10} 160        & \cellcolor{blue!6} 342       & \cellcolor{blue!6} 258       & \cellcolor{blue!6} 141       \\
            + SFT                                                    & \cellcolor{red!10} 35             & \cellcolor{red!10} 97             & \cellcolor{red!10} 50            & \cellcolor{red!6} 37            & \cellcolor{green!6} 19          & \cellcolor{green!6} 56           & \cellcolor{red!10} 96           & \cellcolor{green!10} 54         & \cellcolor{red!10} 215         & \cellcolor{red!6} 272        & \cellcolor{red!6} 235        & \cellcolor{red!6} 114        \\
            + DPO                                                    & \cellcolor{green!10} 4            & \cellcolor{green!10} 12           & \cellcolor{green!10} 10          & \cellcolor{green!6} 26          & \cellcolor{green!6} 76          & \cellcolor{green!6} 30           & \cellcolor{green!10} 41         & \cellcolor{green!10} 40         & \cellcolor{green!10} 120       & \cellcolor{green!6} 369      & \cellcolor{green!6} 277      & \cellcolor{green!6} 275      \\ \bottomrule
            \end{tabular}
    }
\end{table}

\vspace{-0.5em}
\section{Related Work}
\vspace{-0.5em}
\textbf{NL2Code Generation} is the task of generating a computer program that satisfies given specifications. Initial approaches to converting natural language to code relied on rigid methods like probabilistic grammars and domain-specific languages, having limited flexibility and scalability~\cite{joshi2003formalism, de2008z3}. The advent of statistical models, such as n-grams and Hidden Markov models, attempted to overcome these limitations but struggled with modeling complexity and dependencies~\cite{nguyen2013statistical, sutskever2008recurrent}. The transformational impact of the Transformer model~\cite{vaswani2017attention} and its subsequent application to NL2Code~\cite{mastropaolo2021studying} led to the development of LLMs like Codex, which significantly improved the task's feasibility by utilizing extensive unlabelled data sets~\cite{chen2021evaluating}. Follow-up LLMs such as AlphaCode~\cite{li2022competition}, CodeGen~\cite{nijkamp2022codegen}, PaLM-Coder~\cite{chowdhery2022palm}, and StarCoder~\cite{li2023starcoder} continued to advance this research field, exhibiting emergent abilities in coding and debugging that mirrored human programmers.

\textbf{NL2Code Evaluation} currently focuses on gauging the functional correctness of generated code. 
As a pioneer, CodeBLEU~\cite{ren2020codebleu} adapts the BLEU~\cite{papineni2002bleu} metric into code generation. However, given the abstract nature of programming languages, distinct code can express the equivalent semantics, prompting subsequent benchmarks to harness test case fuzzing instead of the similarity measurement. For example, HumanEval~\cite{chen2021evaluating} and MBPP~\cite{austin2021program} consist of hand-written Python programming tasks and corresponding test cases. On the note of enhancing language inclusiveness, ODEX~\cite{wang2022execution} integrates multiple natural languages, while MBXP~\cite{athiwaratkun2022multi} extends the benchmarks to cater to a variety of programming languages, promoting polyglot code generation evaluation. Recent benchmarks have also begun to consider more aspects beyond functional correctness. For instance, the benchmark DS-100~\cite{lai2023ds} dives deeply into the data analysis scenarios, and CodeGen~\cite{nijkamp2022codegen} contributes a benchmark for multi-turn code generation. For security-oriented code generation, SecurityEval~\cite{siddiq2022securityeval} offers a concentrating benchmark on mining the vulnerability of generated code.

\section{Limitations}
\label{sec:limitations}
In this work, we measure code efficiency under the assumption that the code runtime is uniformly distributed. The simplification streamlines code efficiency evaluation via limited solution samples. However, the distribution of code runtime in real-world scenarios is more intricate, which may call for more solution samples to support more precise modeling. Additionally, the presence of data contamination during the model training phase compromises the precision of the Mercury benchmark to reflect the performance of tainted models~\cite{jain2024livecodebench}. To mitigate this issue, we will update our benchmark via our open-sourced data collection framework to import new tasks dynamically, thus laying the groundwork for more detailed investigations in subsequent studies.

\section{Conclusion}
In this work, we introduced Mercury, the first code efficiency benchmark for NL2Code evaluation. Unlike prior work that focused on functional correctness, our benchmark highlights the importance of code efficiency. 
By crafting dedicated test case generators and sampling ground-truth solutions across all difficulty levels from Leetcode, we have developed a comprehensive and rigorous Code LLM evaluation frame. 
We evaluated leading Code LLMs against benchmarks and found that even though these models are proficient in generating functionally correct code, there is still considerable space for code efficiency improvement. 
As Code LLMs become more widely used, code efficiency determines factual productivity, where Mercury can gauge the vital metric. As a commitment to ongoing research and to foster further innovation in this area, we have open-sourced the Mercury dataset collection framework, laying the groundwork for future advancements in the field.

\bibliography{custom}

\begin{thebibliography}{10}

\bibitem{aminabadi2022deepspeed}
Reza~Yazdani Aminabadi, Samyam Rajbhandari, Ammar~Ahmad Awan, Cheng Li, Du~Li, Elton Zheng, Olatunji Ruwase, Shaden Smith, Minjia Zhang, Jeff Rasley, et~al.
\newblock Deepspeed-inference: enabling efficient inference of transformer models at unprecedented scale.
\newblock In {\em SC22: International Conference for High Performance Computing, Networking, Storage and Analysis}, pages 1--15. IEEE, 2022.

\bibitem{athiwaratkun2022multi}
Ben Athiwaratkun, Sanjay~Krishna Gouda, Zijian Wang, Xiaopeng Li, Yuchen Tian, Ming Tan, Wasi~Uddin Ahmad, Shiqi Wang, Qing Sun, Mingyue Shang, et~al.
\newblock Multi-lingual evaluation of code generation models.
\newblock {\em arXiv preprint arXiv:2210.14868}, 2022.

\bibitem{austin2021program}
Jacob Austin, Augustus Odena, Maxwell Nye, Maarten Bosma, Henryk Michalewski, David Dohan, Ellen Jiang, Carrie Cai, Michael Terry, Quoc Le, et~al.
\newblock Program synthesis with large language models.
\newblock {\em arXiv preprint arXiv:2108.07732}, 2021.

\bibitem{qwen}
Jinze Bai, Shuai Bai, Yunfei Chu, Zeyu Cui, Kai Dang, Xiaodong Deng, Yang Fan, Wenbin Ge, Yu~Han, Fei Huang, Binyuan Hui, Luo Ji, Mei Li, Junyang Lin, Runji Lin, Dayiheng Liu, Gao Liu, Chengqiang Lu, Keming Lu, Jianxin Ma, Rui Men, Xingzhang Ren, Xuancheng Ren, Chuanqi Tan, Sinan Tan, Jianhong Tu, Peng Wang, Shijie Wang, Wei Wang, Shengguang Wu, Benfeng Xu, Jin Xu, An~Yang, Hao Yang, Jian Yang, Shusheng Yang, Yang Yao, Bowen Yu, Hongyi Yuan, Zheng Yuan, Jianwei Zhang, Xingxuan Zhang, Yichang Zhang, Zhenru Zhang, Chang Zhou, Jingren Zhou, Xiaohuan Zhou, and Tianhang Zhu.
\newblock Qwen technical report.
\newblock {\em arXiv preprint arXiv:2309.16609}, 2023.

\bibitem{bai2022training}
Yuntao Bai, Andy Jones, Kamal Ndousse, Amanda Askell, Anna Chen, Nova DasSarma, Dawn Drain, Stanislav Fort, Deep Ganguli, Tom Henighan, et~al.
\newblock Training a helpful and harmless assistant with reinforcement learning from human feedback.
\newblock {\em arXiv preprint arXiv:2204.05862}, 2022.

\bibitem{bakker2022fine}
Michiel Bakker, Martin Chadwick, Hannah Sheahan, Michael Tessler, Lucy Campbell-Gillingham, Jan Balaguer, Nat McAleese, Amelia Glaese, John Aslanides, Matt Botvinick, et~al.
\newblock Fine-tuning language models to find agreement among humans with diverse preferences.
\newblock {\em Advances in Neural Information Processing Systems}, 35:38176--38189, 2022.

\bibitem{chen2022learning}
Binghong Chen, Daniel Tarlow, Kevin Swersky, Martin Maas, Pablo Heiber, Ashish Naik, Milad Hashemi, and Parthasarathy Ranganathan.
\newblock Learning to improve code efficiency.
\newblock {\em arXiv preprint arXiv:2208.05297}, 2022.

\bibitem{chen2021evaluating}
Mark Chen, Jerry Tworek, Heewoo Jun, Qiming Yuan, Henrique Ponde de~Oliveira Pinto, Jared Kaplan, Harri Edwards, Yuri Burda, Nicholas Joseph, Greg Brockman, et~al.
\newblock Evaluating large language models trained on code.
\newblock {\em arXiv preprint arXiv:2107.03374}, 2021.

\bibitem{chowdhery2022palm}
A~Chowdhery, S~Narang, J~Devlin, M~Bosma, G~Mishra, A~Roberts, P~Barham, HW~Chung, C~Sutton, S~Gehrmann, et~al.
\newblock Palm: Scaling language modeling with pathways (no. arxiv: 2204.02311). arxiv, 2022.

\bibitem{cc_by_nc_4}
Creative Commons.
\newblock Cc by-nc 4.0 deed.
\newblock \url{https://creativecommons.org/licenses/by-nc/4.0/}, 2024.
\newblock [Accessed 25-05-2024].

\bibitem{de2008z3}
Leonardo De~Moura and Nikolaj Bj{\o}rner.
\newblock Z3: An efficient smt solver.
\newblock In {\em International conference on Tools and Algorithms for the Construction and Analysis of Systems}, pages 337--340. Springer, 2008.

\bibitem{deepseekcoder}
Deepseek-Ai.
\newblock Deepseek-ai/deepseek-coder: Deepseek coder: Let the code write itself, 2023.

\bibitem{dettmers2022llmint8}
Tim Dettmers, Mike Lewis, Younes Belkada, and Luke Zettlemoyer.
\newblock Llm.int8(): 8-bit matrix multiplication for transformers at scale.
\newblock {\em arXiv preprint arXiv:2208.07339}, 2022.

\bibitem{accelerate}
Sylvain Gugger, Lysandre Debut, Thomas Wolf, Philipp Schmid, Zachary Mueller, Sourab Mangrulkar, Marc Sun, and Benjamin Bossan.
\newblock Accelerate: Training and inference at scale made simple, efficient and adaptable.
\newblock \url{https://github.com/huggingface/accelerate}, 2022.

\bibitem{hendrycks2021measuring}
Dan Hendrycks, Steven Basart, Saurav Kadavath, Mantas Mazeika, Akul Arora, Ethan Guo, Collin Burns, Samir Puranik, Horace He, Dawn Song, et~al.
\newblock Measuring coding challenge competence with apps.
\newblock {\em arXiv preprint arXiv:2105.09938}, 2021.

\bibitem{hu2021lora}
Edward~J Hu, Yelong Shen, Phillip Wallis, Zeyuan Allen-Zhu, Yuanzhi Li, Shean Wang, Lu~Wang, and Weizhu Chen.
\newblock Lora: Low-rank adaptation of large language models.
\newblock {\em arXiv preprint arXiv:2106.09685}, 2021.

\bibitem{jafari2021survey}
Omid Jafari, Preeti Maurya, Parth Nagarkar, Khandker~Mushfiqul Islam, and Chidambaram Crushev.
\newblock A survey on locality sensitive hashing algorithms and their applications.
\newblock {\em arXiv preprint arXiv:2102.08942}, 2021.

\bibitem{jain2024livecodebench}
Naman Jain, King Han, Alex Gu, Wen-Ding Li, Fanjia Yan, Tianjun Zhang, Sida Wang, Armando Solar-Lezama, Koushik Sen, and Ion Stoica.
\newblock Livecodebench: Holistic and contamination free evaluation of large language models for code.
\newblock {\em arXiv preprint arXiv:2403.07974}, 2024.

\bibitem{joshi2003formalism}
Aravind Joshi and Owen Rambow.
\newblock A formalism for dependency grammar based on tree adjoining grammar.
\newblock In {\em Proceedings of the Conference on Meaning-text Theory}, pages 207--216. MTT Paris, France, 2003.

\bibitem{kir2017overcome}
James Kirkpatrick, Razvan Pascanu, Neil Rabinowitz, Joel Veness, Guillaume Desjardins, Andrei~A. Rusu, Kieran Milan, John Quan, Tiago Ramalho, Agnieszka Grabska-Barwinska, Demis Hassabis, Claudia Clopath, Dharshan Kumaran, and Raia Hadsell.
\newblock Overcoming catastrophic forgetting in neural networks.
\newblock {\em Proceedings of the National Academy of Sciences - PNAS}, 114(13):3521--3526, 2017.

\bibitem{kulal2019spoc}
Sumith Kulal, Panupong Pasupat, Kartik Chandra, Mina Lee, Oded Padon, Alex Aiken, and Percy~S Liang.
\newblock Spoc: Search-based pseudocode to code.
\newblock {\em Advances in Neural Information Processing Systems}, 32, 2019.

\bibitem{lai2023ds}
Yuhang Lai, Chengxi Li, Yiming Wang, Tianyi Zhang, Ruiqi Zhong, Luke Zettlemoyer, Wen-tau Yih, Daniel Fried, Sida Wang, and Tao Yu.
\newblock Ds-1000: A natural and reliable benchmark for data science code generation.
\newblock In {\em International Conference on Machine Learning}, pages 18319--18345. PMLR, 2023.

\bibitem{leetcode}
LeetCode.
\newblock {L}eet{C}ode.
\newblock \url{https://leetcode.com/problemset/algorithms/}, 2024.
\newblock [Accessed 25-05-2024].

\bibitem{li2023starcoder}
Raymond Li, Loubna~Ben Allal, Yangtian Zi, Niklas Muennighoff, Denis Kocetkov, Chenghao Mou, Marc Marone, Christopher Akiki, Jia Li, Jenny Chim, et~al.
\newblock Starcoder: may the source be with you!
\newblock {\em arXiv preprint arXiv:2305.06161}, 2023.

\bibitem{li2022competition}
Yujia Li, David Choi, Junyoung Chung, Nate Kushman, Julian Schrittwieser, R{\'e}mi Leblond, Tom Eccles, James Keeling, Felix Gimeno, Agustin Dal~Lago, et~al.
\newblock Competition-level code generation with alphacode.
\newblock {\em Science}, 378(6624):1092--1097, 2022.

\bibitem{evalplusliu}
Jiawei Liu, Chunqiu~Steven Xia, Yuyao Wang, and Lingming Zhang.
\newblock Is your code generated by chatgpt really correct?
\newblock {\em arXiv}, 2023.

\bibitem{liu2024your}
Jiawei Liu, Chunqiu~Steven Xia, Yuyao Wang, and Lingming Zhang.
\newblock Is your code generated by chatgpt really correct? rigorous evaluation of large language models for code generation.
\newblock {\em Advances in Neural Information Processing Systems}, 36, 2024.

\bibitem{loshchilov2017decoupled}
Ilya Loshchilov and Frank Hutter.
\newblock Decoupled weight decay regularization.
\newblock {\em arXiv preprint arXiv:1711.05101}, 2017.

\bibitem{lozhkov2024starcoder}
Anton Lozhkov, Raymond Li, Loubna~Ben Allal, Federico Cassano, Joel Lamy-Poirier, Nouamane Tazi, Ao~Tang, Dmytro Pykhtar, Jiawei Liu, Yuxiang Wei, et~al.
\newblock Starcoder 2 and the stack v2: The next generation.
\newblock {\em arXiv preprint arXiv:2402.19173}, 2024.

\bibitem{mastropaolo2021studying}
Antonio Mastropaolo, Simone Scalabrino, Nathan Cooper, David~Nader Palacio, Denys Poshyvanyk, Rocco Oliveto, and Gabriele Bavota.
\newblock Studying the usage of text-to-text transfer transformer to support code-related tasks.
\newblock In {\em 2021 IEEE/ACM 43rd International Conference on Software Engineering (ICSE)}, pages 336--347. IEEE, 2021.

\bibitem{nguyen2013statistical}
Tung~Thanh Nguyen, Anh~Tuan Nguyen, Hoan~Anh Nguyen, and Tien~N Nguyen.
\newblock A statistical semantic language model for source code.
\newblock In {\em Proceedings of the 2013 9th Joint Meeting on Foundations of Software Engineering}, pages 532--542, 2013.

\bibitem{nijkamp2022codegen}
Erik Nijkamp, Bo~Pang, Hiroaki Hayashi, Lifu Tu, Huan Wang, Yingbo Zhou, Silvio Savarese, and Caiming Xiong.
\newblock Codegen: An open large language model for code with multi-turn program synthesis.
\newblock {\em arXiv preprint arXiv:2203.13474}, 2022.

\bibitem{fairuse}
U.S.~Copyright Office.
\newblock U.s. copyright office fair use index.
\newblock \url{https://www.copyright.gov/fair-use/}, 2024.
\newblock [Accessed 25-05-2024].

\bibitem{openai2023gpt4}
R~OpenAI.
\newblock Gpt-4 technical report.
\newblock {\em arXiv}, pages 2303--08774, 2023.

\bibitem{papineni2002bleu}
Kishore Papineni, Salim Roukos, Todd Ward, and Wei-Jing Zhu.
\newblock Bleu: a method for automatic evaluation of machine translation.
\newblock In {\em Proceedings of the 40th annual meeting of the Association for Computational Linguistics}, pages 311--318, 2002.

\bibitem{rafailov2023direct}
Rafael Rafailov, Archit Sharma, Eric Mitchell, Stefano Ermon, Christopher~D Manning, and Chelsea Finn.
\newblock Direct preference optimization: Your language model is secretly a reward model.
\newblock {\em arXiv preprint arXiv:2305.18290}, 2023.

\bibitem{ren2020codebleu}
Shuo Ren, Daya Guo, Shuai Lu, Long Zhou, Shujie Liu, Duyu Tang, Neel Sundaresan, Ming Zhou, Ambrosio Blanco, and Shuai Ma.
\newblock Codebleu: a method for automatic evaluation of code synthesis.
\newblock {\em arXiv preprint arXiv:2009.10297}, 2020.

\bibitem{roziere2023code}
Baptiste Roziere, Jonas Gehring, Fabian Gloeckle, Sten Sootla, Itai Gat, Xiaoqing~Ellen Tan, Yossi Adi, Jingyu Liu, Tal Remez, J{\'e}r{\'e}my Rapin, et~al.
\newblock Code llama: Open foundation models for code.
\newblock {\em arXiv preprint arXiv:2308.12950}, 2023.

\bibitem{siddiq2022securityeval}
Mohammed~Latif Siddiq and Joanna~CS Santos.
\newblock Securityeval dataset: mining vulnerability examples to evaluate machine learning-based code generation techniques.
\newblock In {\em Proceedings of the 1st International Workshop on Mining Software Repositories Applications for Privacy and Security}, pages 29--33, 2022.

\bibitem{stiennon2020learning}
Nisan Stiennon, Long Ouyang, Jeffrey Wu, Daniel Ziegler, Ryan Lowe, Chelsea Voss, Alec Radford, Dario Amodei, and Paul~F Christiano.
\newblock Learning to summarize with human feedback.
\newblock {\em Advances in Neural Information Processing Systems}, 33:3008--3021, 2020.

\bibitem{sutskever2008recurrent}
Ilya Sutskever, Geoffrey~E Hinton, and Graham~W Taylor.
\newblock The recurrent temporal restricted boltzmann machine.
\newblock {\em Advances in neural information processing systems}, 21, 2008.

\bibitem{vaswani2017attention}
Ashish Vaswani, Noam Shazeer, Niki Parmar, Jakob Uszkoreit, Llion Jones, Aidan~N Gomez, {\L}ukasz Kaiser, and Illia Polosukhin.
\newblock Attention is all you need.
\newblock {\em Advances in neural information processing systems}, 30, 2017.

\bibitem{wang2022execution}
Zhiruo Wang, Shuyan Zhou, Daniel Fried, and Graham Neubig.
\newblock Execution-based evaluation for open-domain code generation.
\newblock {\em arXiv preprint arXiv:2212.10481}, 2022.

\bibitem{wong2023natural}
Man-Fai Wong, Shangxin Guo, Ching-Nam Hang, Siu-Wai Ho, and Chee-Wei Tan.
\newblock Natural language generation and understanding of big code for ai-assisted programming: A review.
\newblock {\em Entropy}, 25(6):888, 2023.

\bibitem{xu2022survey}
Yichen Xu and Yanqiao Zhu.
\newblock A survey on pretrained language models for neural code intelligence.
\newblock {\em arXiv preprint arXiv:2212.10079}, 2022.

\bibitem{zan2022large}
Daoguang Zan, Bei Chen, Fengji Zhang, Dianjie Lu, Bingchao Wu, Bei Guan, Yongji Wang, and Jian-Guang Lou.
\newblock Large language models meet nl2code: A survey.
\newblock {\em arXiv preprint arXiv:2212.09420}, 2022.

\bibitem{zan2023large}
Daoguang Zan, Bei Chen, Fengji Zhang, Dianjie Lu, Bingchao Wu, Bei Guan, Wang Yongji, and Jian-Guang Lou.
\newblock Large language models meet nl2code: A survey.
\newblock In {\em Proceedings of the 61st Annual Meeting of the Association for Computational Linguistics (Volume 1: Long Papers)}, pages 7443--7464, 2023.

\bibitem{ziegler2019fine}
Daniel~M Ziegler, Nisan Stiennon, Jeffrey Wu, Tom~B Brown, Alec Radford, Dario Amodei, Paul Christiano, and Geoffrey Irving.
\newblock Fine-tuning language models from human preferences.
\newblock {\em arXiv preprint arXiv:1909.08593}, 2019.

\end{thebibliography}
\bibliographystyle{plain}

\clearpage
\section*{Checklist}


\begin{enumerate}

\item For all authors...
\begin{enumerate}
  \item Do the main claims made in the abstract and introduction accurately reflect the paper's contributions and scope?
    \answerYes{The abstract and introduction include the main contributions and the research scope.} 
  \item Did you describe the limitations of your work?
    \answerYes{See Section~\ref{sec:limitations}.}
  \item Did you discuss any potential negative societal impacts of your work?
    \answerNA{Given that our work is largely technical and does not engage with societal systems directly, it is unlikely to have negative societal repercussions.}
  \item Have you read the ethics review guidelines and ensured that your paper conforms to them?
    \answerYes{The paper conforms to the ethics review guidelines.}
\end{enumerate}

\item If you are including theoretical results...
\begin{enumerate}
  \item Did you state the full set of assumptions of all theoretical results?
    \answerNA{The paper does not include any theoretical results.}
  \item Did you include complete proofs of all theoretical results?
    \answerNA{The paper does not include any theoretical results.}
\end{enumerate}

\item If you ran experiments (e.g. for benchmarks)...
\begin{enumerate}
  \item Did you include the code, data, and instructions needed to reproduce the main experimental results (either in the supplemental material or as a URL)?
    \answerYes{We release our dataset on HuggingFace and our code on GitHub. See the abstract footnote.}
  \item Did you specify all the training details (e.g., data splits, hyperparameters, how they were chosen)?
    \answerYes{See Section~\ref{sec:experimental_setups}.}
	\item Did you report error bars (e.g., with respect to the random seed after running experiments multiple times)?
    \answerYes{See Section~\ref{sec:experiments}.}
	\item Did you include the total amount of compute and the type of resources used (e.g., type of GPUs, internal cluster, or cloud provider)?
    \answerYes{See Section~\ref{sec:experimental_setups} and Appendix Section~\ref{sec:score_distribution}.}
\end{enumerate}

\item If you are using existing assets (e.g., code, data, models) or curating/releasing new assets...
\begin{enumerate}
  \item If your work uses existing assets, did you cite the creators?
    \answerYes{See References.}
  \item Did you mention the license of the assets?
    \answerYes{See Section~\ref{sec:legal_compliance}}
  \item Did you include any new assets either in the supplemental material or as a URL?
    \answerYes{See the abstract footnote.}
  \item Did you discuss whether and how consent was obtained from people whose data you're using/curating?
    \answerYes{See Section~\ref{sec:legal_compliance}}
  \item Did you discuss whether the data you are using/curating contains personally identifiable information or offensive content?
    \answerYes{See Section~\ref{sec:legal_compliance}}
\end{enumerate}

\item If you used crowdsourcing or conducted research with human subjects...
\begin{enumerate}
  \item Did you include the full text of instructions given to participants and screenshots, if applicable?
    \answerNA{}
  \item Did you describe any potential participant risks, with links to Institutional Review Board (IRB) approvals, if applicable?
    \answerNA{}
  \item Did you include the estimated hourly wage paid to participants and the total amount spent on participant compensation?
    \answerNA{}
\end{enumerate}

\end{enumerate}


\newpage
\appendix

\section{Appendix}
\label{sec:appendix}

\subsection{Dataset Nutrition Labels}
\begin{table*}[h!]
    \caption{Definitions of the fields within the Mercury dataset.}
    \label{tab:mercury_fields}
    \resizebox{\textwidth}{!}{
    \begin{tabular}{ll}
        \toprule
        \textbf{Field Name} & \textbf{Definition}                                                                   \\ \midrule
        id                  & Task ID.                                                                              \\
        slug\_name          & Task name.                                                                            \\
        meta\_info          & The field accommodating the task description and submission statistics.               \\
        difficulty          & The difficulty level of the task.                                                     \\
        pretty\_content     & The field introduces the task description, examples, and constraints in pure text.    \\
        solutions           & Samples of solutions extracted from actual past submissions.                          \\
        prompt              & The prompt of the solution.                                                           \\
        entry\_point        & The nominative entry point of the solution.                                           \\
        generator\_code     & A function to generate test cases.                                                    \\
        test\_cases         & A collection of generated test cases.                                                 \\
        convert\_online     & A function to format test cases for online evaluation.                                \\
        convert\_offline    & A function to format test cases for offline evaluation.                               \\
        evaluate\_offline   & A function designed to evaluate solutions in an offline setting.                      \\ 
        \bottomrule
    \end{tabular}
    }
\end{table*}

\subsection{Mercury Data Distribution and Customized Data Structures }
\begin{figure}[h!]
    \centering
    \begin{minipage}{0.5\linewidth}
        \centering
        \includegraphics[width=\linewidth]{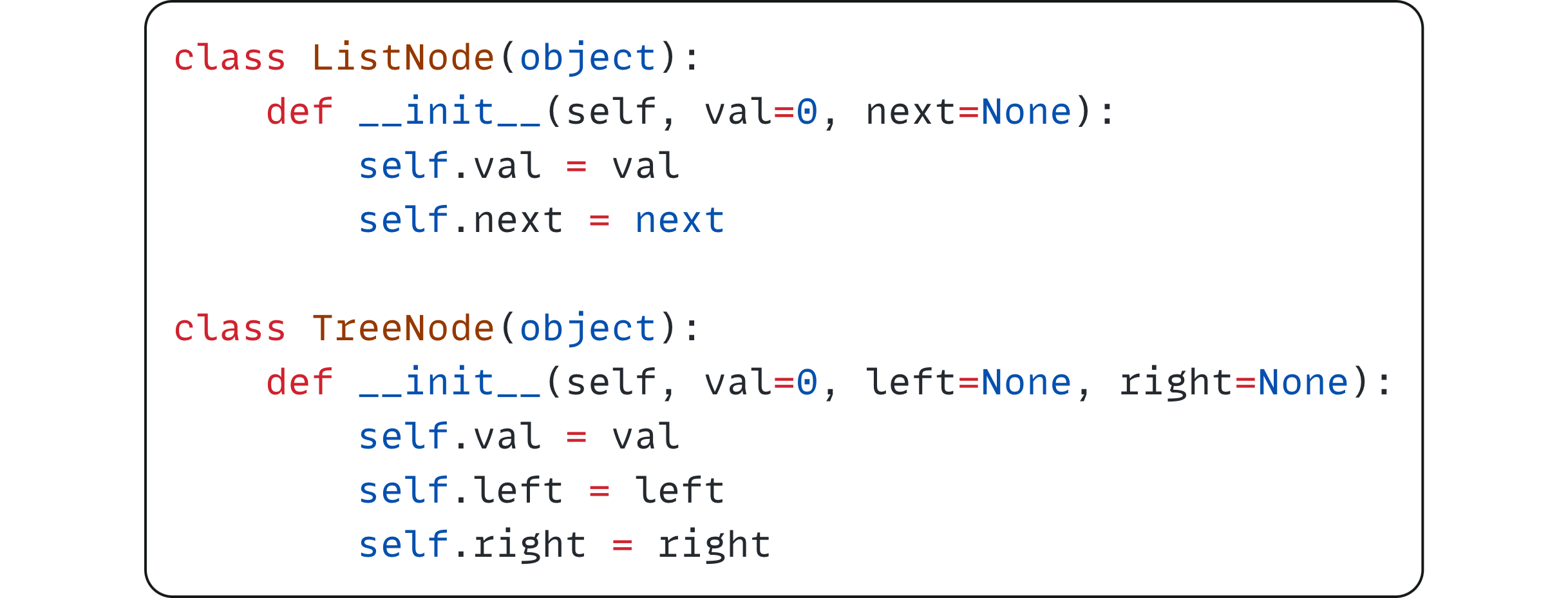}
        \caption{Mercury supports two customized data structures: TreeNode and ListNode.}
        \label{fig:mercury_data_structure}
    \end{minipage}
    \hfill 
    \begin{minipage}{0.45\linewidth}
        \centering
        \resizebox{\linewidth}{!}{
            \begin{tabular}{lcccc}
                \toprule
                \textbf{Splits}     & \textbf{Easy} & \textbf{Medium} & \textbf{Hard} & \textbf{Sum} \\ 
                \midrule
                Mercury-train       & 446           & 968             & 219           & 1,633        \\
                Mercury-eval        & 88            & 81              & 87            & 256          \\ 
                \bottomrule
            \end{tabular}
        }
        \captionof{table}{\emph{Mercury-eval} encompasses 256 tasks, the difficulty level of which has been balanced for model evaluation. \emph{Mercury-train} comprises the remaining 1,633 tasks for model training.}
        \label{tab:dataset_distribution}
    \end{minipage}
\end{figure}

Except for all built-in Python data structures, Mercury imports another two structures to enhance the diversity and complexity as shown in Figure~\ref{fig:mercury_data_structure}.


\subsection{Sandbox Details}
\label{sec:sandbox_details}
\paragraph{Time and Memory Limitation.} 
Each executed code within the sandbox is subject to certain constraints to ensure fair utilization of resources and to prevent any single code from monopolizing the system resource. Specifically, there are two primary constraints: a time limit and a memory limit. The time limit restricts how long the code can execute before being forcibly terminated, thereby ensuring that no infinite loops or excessively long computations negatively impact the availability of the sandbox. The memory limit caps the amount of RAM that a process can consume. This measure precludes a single code from exhausting the memory resources, which could lead to a denial of service for subsequent codes. In our experiment settings, the timeout limit is 30 seconds, and the memory limit is 2048 MB for each solution execution.

\paragraph{IO Restriction.}
To mitigate harmful activities such as unauthorized command execution or data exfiltration, the sandbox imposes strict Input/Output (IO) restrictions. These restrictions include limitations on reading from or writing to the disk and restrictions on the use of network sockets for sending or receiving data. By controlling the IO operations, the sandbox can prevent many common vulnerabilities and ensure that the code runs without interfering with other processes of the host system.

\paragraph{Isolated File System.}
The sandbox employs an isolated file system to provide a safe execution environment for the code. This means that the process running in the sandbox has its virtual file system, which is separated from the host's file system. The isolated nature of this file system ensures that even if a process within the sandbox attempts to modify or delete files, these changes will not affect the host system or other sandboxes. It acts as a security layer, protecting the host from potential threats and maintaining the integrity of the overall system.

\paragraph{System Libraries Redirection.}
To maintain a consistent and controlled environment, the sandbox redirects calls to system libraries to sandbox-specific versions. This is done to prevent code from using certain functions directly from the host's system libraries, which could result in unpredictable behavior or security vulnerabilities. The redirected libraries are often limited to a subset of functionalities deemed safe and necessary for executing programs within the sandbox, thus enforcing the security policies and ensuring that the running programs behave as expected.

\paragraph{Single-threaded Evaluation.}
Single-threaded evaluation refers to executing code using a sole thread of execution, thereby simplifying resource management and timing assessments, and mitigating the intricacies linked with multi-threaded execution, such as synchronization issues, race conditions, and potential deadlocks. This mode of operation is especially important in testing environments where reproducibility and fairness are paramount, ensuring that each piece of code is evaluated using identical computational resources.

\begin{figure}[ht]
    \centering
    \includegraphics[width=0.95\linewidth]{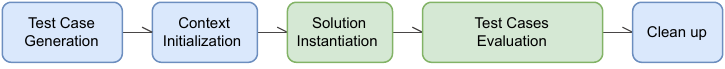}
    \caption{
        Sandbox Execution Pipeline. 
        \textbf{1) Test Case Generation.} We first employ the corresponding test case generator for each task to produce a comprehensive set of test cases for the subsequent evaluation.
        \textbf{2) Context Initialization.} To prevent any unexpected code behavior, the sandbox environment is meticulously reinitialized for each new task. This phase ensures that all the common libraries required for executing the solution are loaded.
        \textbf{3) Solution Instantiation.} The solution under evaluation will be encapsulated as a \emph{solution} class. 
        \textbf{4) Test Case Evaluation.} Each test case the generator provides will be rigorously executed against the solution. A solution must successfully pass all the test cases to be deemed valid.
        \textbf{5) Clean up.} The final stage involves the sandbox dutifully clearing the namespace environment and the temporary directory.
        Mercury records the time consumed during the stage of Solution instantiation and Test Ease Evaluation as the primary metric for assessing code efficiency.   
    }
    \label{fig:sandbox_code_efficiency}
\end{figure}

\paragraph{Code Efficiency Measurement.}
Figure~\ref{fig:sandbox_code_efficiency} shows the overview of the code execution pipeline. We gauge the \emph{Solution Instantiation} and \emph{Test Ease Evaluation} time spans as the execution runtime.

\subsection{DPO Experiment Details}
\paragraph{Dataset Construction.} For every task problem $T^{i}$ in Mercury, we randomly selected two solutions from the task solution set $\{s^{i}_w, s^{i}_l\} \sim T^{i}_{solution}$, to construct the preference dataset $D = \{ P^{i}, s^{i}_{w}, s^{i}_{l} \}$, where $p^{i}$ is the prompt, $s^{i}_{w}$ has a faster runtime than $s^{i}_{l}$.

\paragraph{Model Initialization.} RLHF~\cite{ziegler2019fine} typically begins with a reference LLM $\pi_{ref}$. Here, we initialize $\pi_{ref}$ by maximizing the likelihood of faster code completions $(p, s_{w}) \sim D$, so that $\pi_{ref} = \arg \max_{\pi} E_{(p, s_{w}) \sim D} \left[ \log \pi (s_{w} | p) \right] $. This procedure helps mitigate the distribution shift between the \emph{true reference distribution} and $\pi_{ref}$.

\paragraph{Optimization.} We optimize the target LLM $\pi_{\theta}$ to minimize $\mathcal{L}_{DPO}$ for the given $\pi_{ref}$ and $D$ and desired hyperparameter $\beta$. The gradient with respect to the parameters $\theta$ can be written as $\nabla_{\theta} \mathcal{L}_{DPO}(\pi_{\theta};\pi_{ref})$.

\label{sec:dpo_details}
\begin{equation}
    \mathcal{L}_{DPO}(\pi_{\theta};\pi_{ref}) = 
    -E_{(x, s_{w}, s_{l})\sim D} \left[ \log \alpha (\beta \log \frac{\pi_{\theta}(s_w|p)}{\pi_{ref}(s_w|p)}) - \log \frac{\pi_{\theta}(s_l|p)}{\pi_{ref}(s_l|p)}) \right]
\end{equation}

\begin{multline}
    \nabla_{\theta} \mathcal{L}_{DPO}(\pi_{\theta};\pi_{ref}) = \\
    - \beta E_{(p, s_{w}, s_{l})\sim D} \left[ 
        \underbrace{\alpha (\hat{r}_{\theta}(p, s_l) - \hat{r}_{\theta}(p, s_w))}_{\textit{higher weight for wrong estimate}}
        \left[ 
            \underbrace{\nabla_{\theta} \log \pi (s_{w}|p)}_{\textit{increase likelihood of $s_w$} } - \underbrace{\nabla_{\theta} \log \pi (s_{l}|p)}_{\textit{decrease likelihood of $s_l$} } 
        \right] 
    \right]
\end{multline}

Intuitively, the gradient of the loss function $\mathcal{L}_{DPO}$ increases the likelihood of the preferred completions $s_w$ and decreases the likelihood of dis-preferred completions $s_l$, which are weighed by how much higher the implicit reward model $\hat{r}_{\theta}$ rates the dis-preferred completions, scaled by $\beta$, \textit{i.e.}, how incorrectly the implicit reward model orders the completions, accounting for the strength of the KL constraint. 

\subsection{External Libraries Utilized in Mercury}
\begin{figure}[ht]
    \centering
    \includegraphics[width=0.75\linewidth]{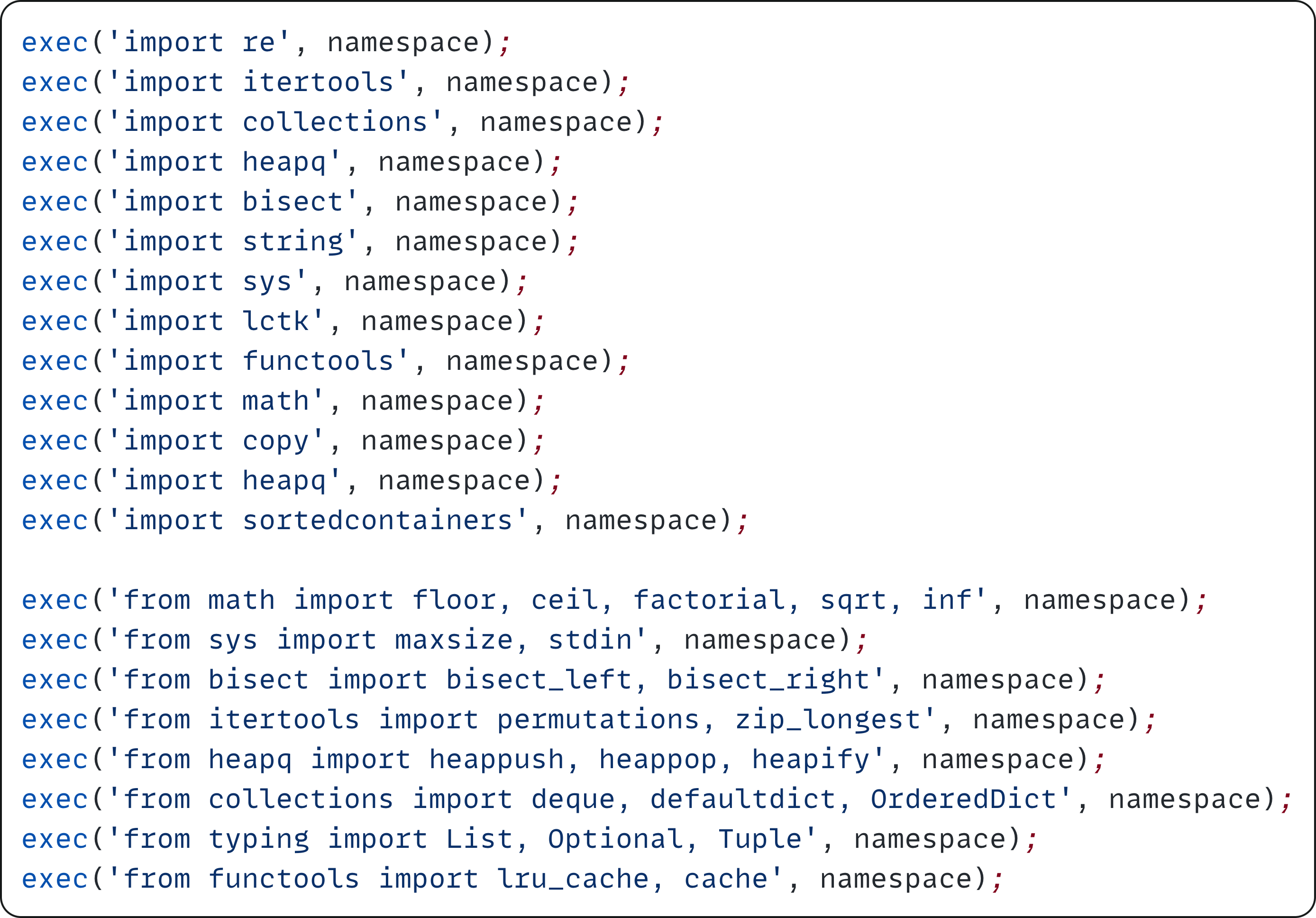}
    \caption{
        External Libraries Imported in Mercury Evaluate Framework.
    }
    \label{fig:external_library}
\end{figure}

Raw LeetCode solutions typically commence without importing shared libraries. To avoid solution failure due to absent libraries, we proactively import the libraries listed in Figure~\ref{fig:external_library} during the sandbox \emph{Context Initialization} phase. Note that all these libraries are imported in a temporary namespace of which the sandbox controls code behaviors. 

\subsection{Model Details}

\begin{table}[ht]
    \caption{Model Details. We evaluated LLMs ranging from 1.3B to 34B.}
    \label{tab:model_details}
    \resizebox{\linewidth}{!}{
        \begin{tabular}{lcl}
            \toprule
            \multicolumn{1}{l}{\textbf{Model Name}} & \multicolumn{1}{l}{\textbf{Model Scale}}      & \multicolumn{1}{l}{\textbf{Link}}                                         \\ \midrule
            deepseek-coder-1.3b-base               & 1.3B                                           & https://huggingface.co/deepseek-ai/deepseek-coder-1.3b-base               \\
            starcoder2-3b                          & 3B                                             & https://huggingface.co/bigcode/starcoder2-3b                              \\
            deepseek-coder-6.7b-base               & 6.7B                                           & https://huggingface.co/deepseek-ai/deepseek-coder-6.7b-base               \\
            starcoder2-7b                          & 7B                                             & https://huggingface.co/bigcode/starcoder2-7b                              \\
            CodeLlama-7b-hf                        & 7B                                             & https://huggingface.co/codellama/CodeLlama-7b-hf                          \\
            CodeQwen1.5-7B                         & 7B                                             & https://huggingface.co/Qwen/CodeQwen1.5-7B                                \\
            CodeLlama-13b-hf                       & 13B                                            & https://huggingface.co/codellama/CodeLlama-13b-hf                         \\
            starcoder2-15b                         & 15B                                            & https://huggingface.co/bigcode/starcoder2-15b                             \\
            deepseek-coder-33b-base                & 33B                                            & https://huggingface.co/deepseek-ai/deepseek-coder-33b-base                \\
            CodeLlama-34b-hf                       & 34B                                            & https://huggingface.co/codellama/CodeLlama-34b-hf                         \\
            \bottomrule
        \end{tabular}
    }
\end{table}

\subsection{A Mercury Example}
\begin{figure*}[!ht]
    \centering
    \includegraphics[width=\linewidth]{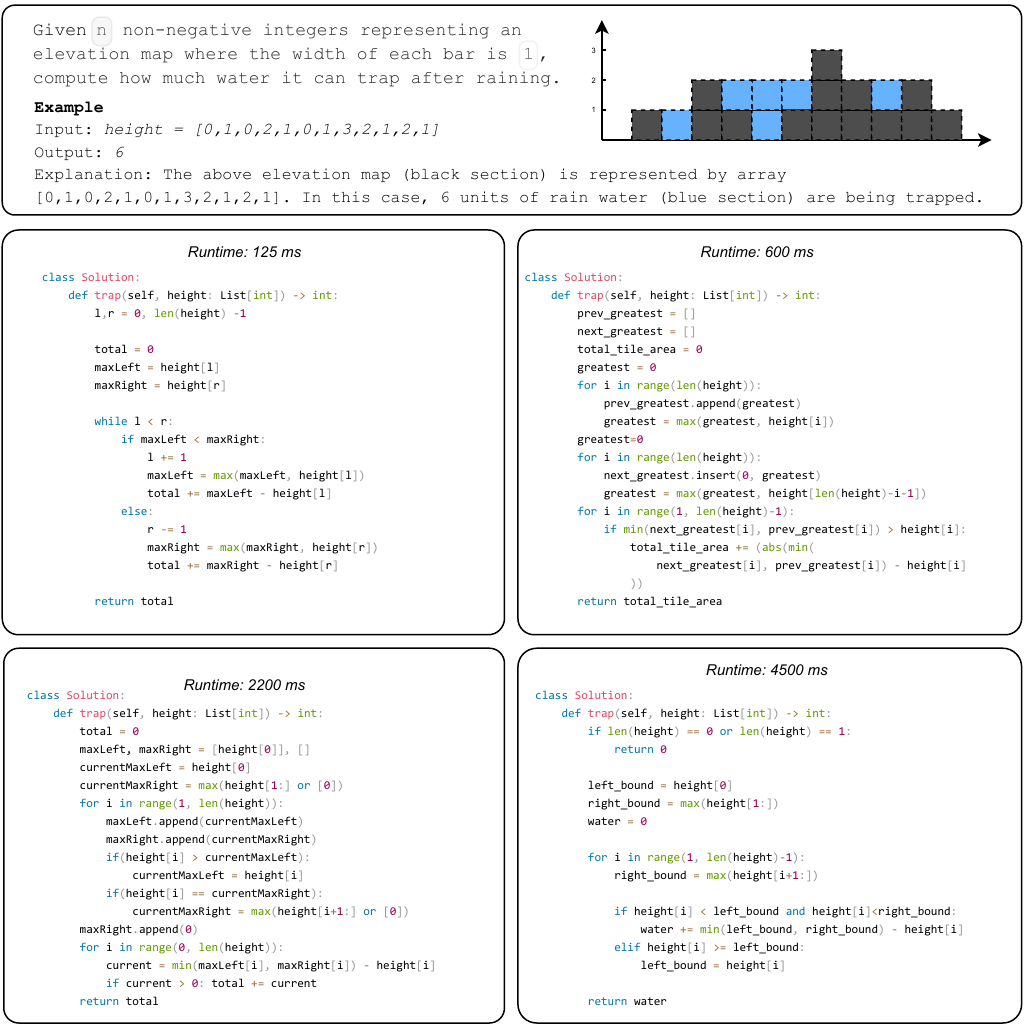}
    \caption{This case is drawn from the \emph{Mercury-eval} benchmark. The upper block presents the problem statement with its example, while the subsequent portion exhibits the corresponding solutions. Although all solutions are functionally correct, they exhibit significant differences in runtimes.}
\end{figure*}

\vspace{-1em}
\subsection{A HumanEval Example}
\begin{figure*}[!ht]
    \centering
    \includegraphics[width=0.9\linewidth]{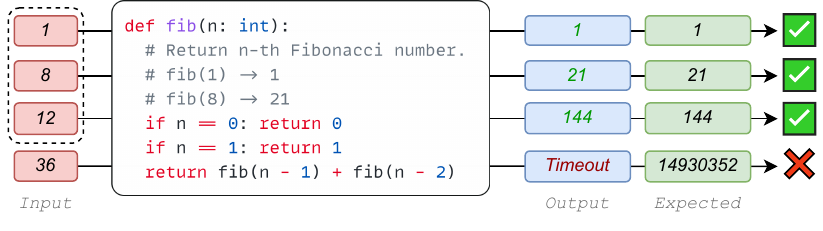}
    \caption{
        An HumanEval example of insufficient test cases. Even though the code passed all test cases in the dashed-line box, it remains vulnerable to timeout or stack overflow when subjected to a larger input.
    }
    \label{fig:mercury_comparison}
\end{figure*}

\subsection{Prompts for Code Generation}
\label{sec:model_prompt}
\begin{figure}[!ht]
    \centering
    \includegraphics[width=0.8\linewidth]{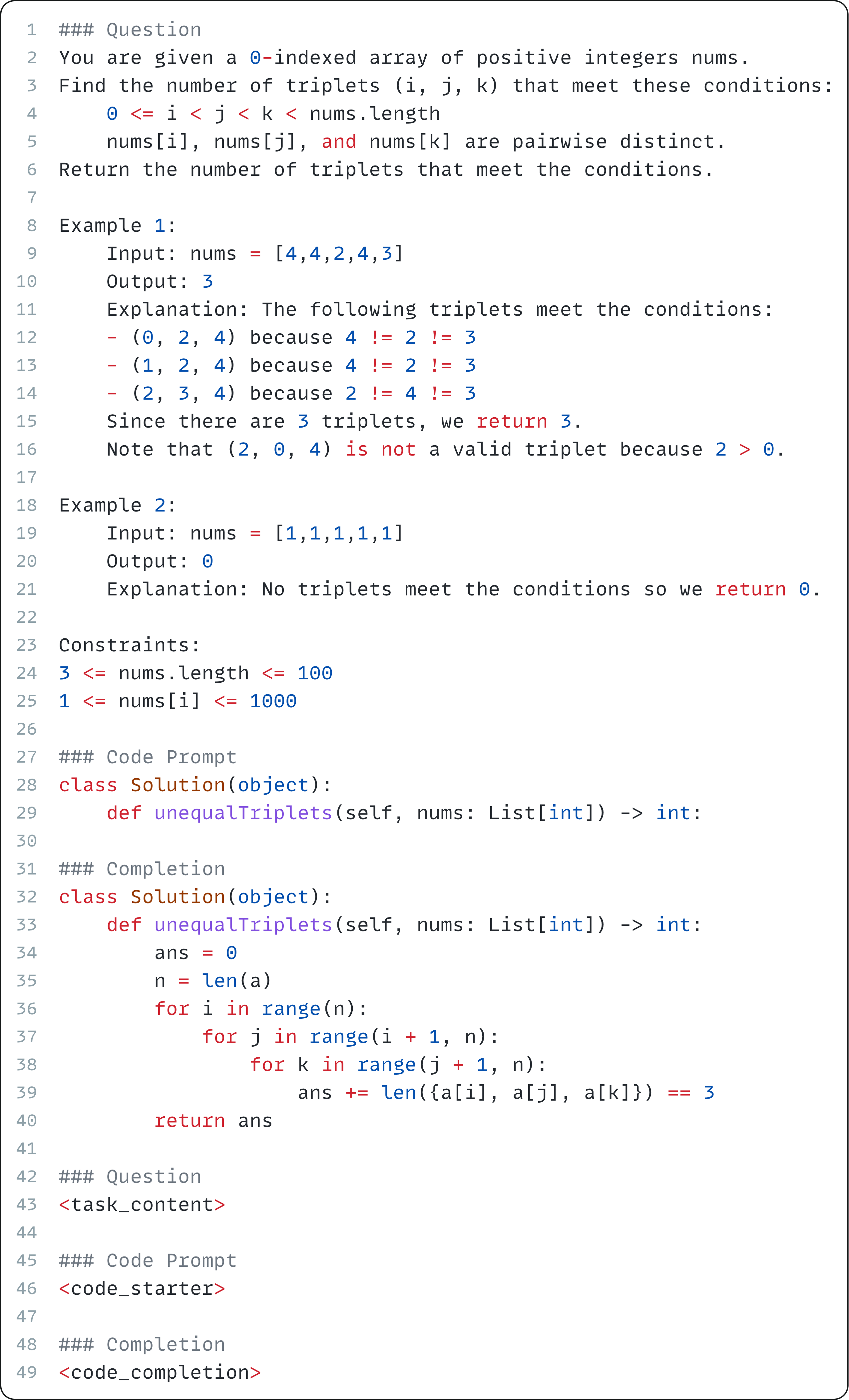}
    \caption{Code Generation Prompts. Lines 1 to 40 are the one-shot example. In Mercury experiments, we feed the \emph{pretty\_content} field to the placeholder \emph{<task\_content>}, the \emph{prompt} field to the placeholder \emph{<code\_starter>}, and the \emph{solution} field to the placeholder \emph{<code\_completion>}}
    \label{fig:model_prompt}
\end{figure}

To guarantee a fair comparison, we apply a unified one-shot prompt template for each pre-trained Code LLM. As displayed in Figure~\ref{fig:model_prompt}, the prompt template contains one shot example as well as three placeholders: \emph{<task\_content>}, \emph{<code\_starter>}, and \emph{<code\_completion>}.

\subsection{Hardware-agnostic Evaluation}
\begin{figure*}[!ht]
    \centering
    \includegraphics[width=0.55\linewidth]{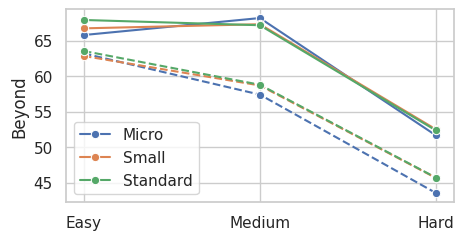}
    \caption{
        \texttt{Beyond} scores of `deepseek-coder-33b'~(solid line) and `deepseek-coder-6.7b'~(dashed line) across varied Intel Skylake CPU configurations. 
        The results show that \texttt{Beyond} can remain consistent across different hardware configurations.
        \vspace{-1em}
    }
    \label{fig:hardware_independent}
\end{figure*}

\subsection{Distribution of Bootstrapped Beyond Scores}
\label{sec:score_distribution}
\begin{figure*}[!ht]
    \centering
    \includegraphics[width=0.6\linewidth]{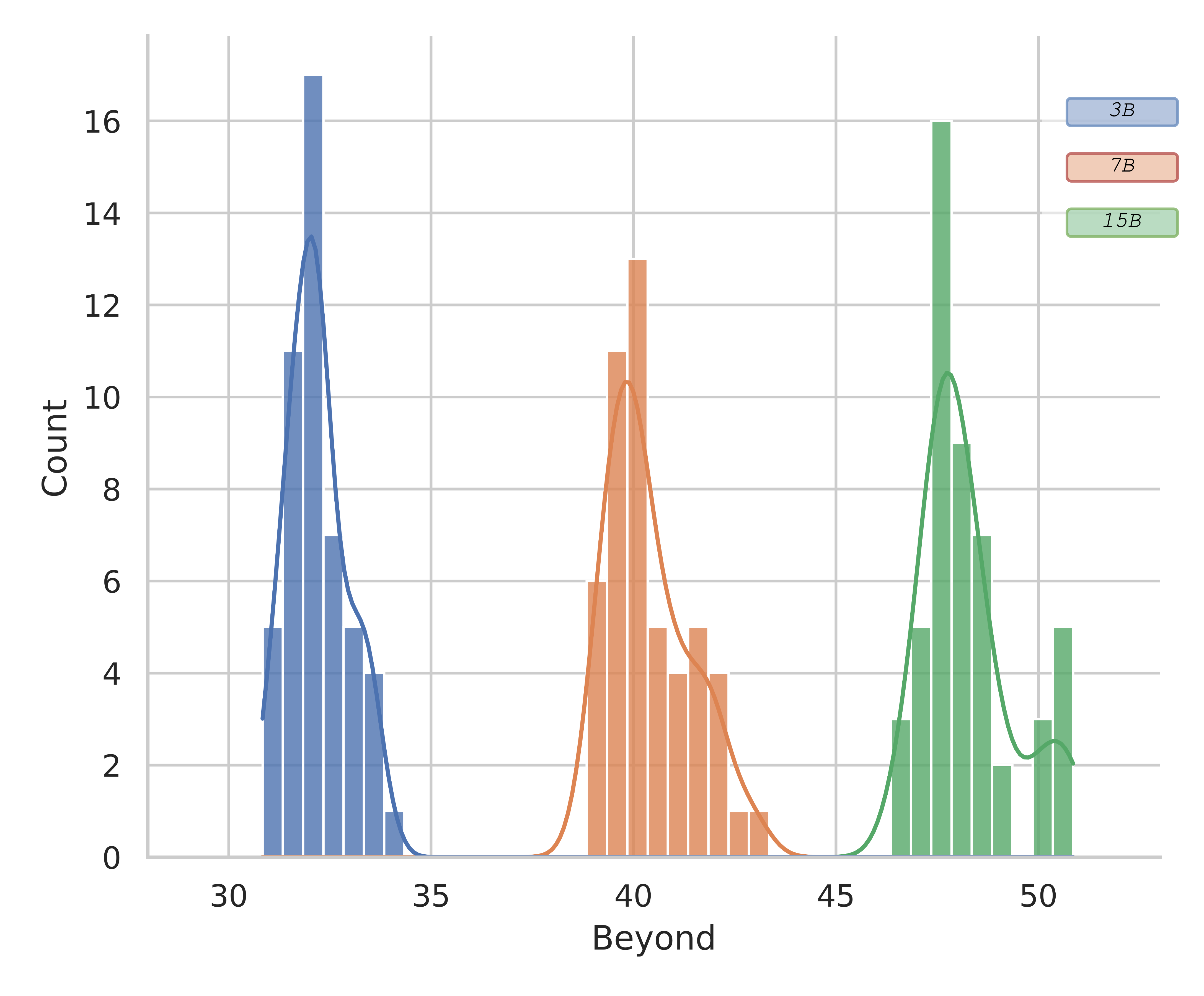}
    \caption{
        Bootstrapped \texttt{Beyond} Distribution. We evaluate 3B, 7B, and 15B Starcoder2\cite{li2023starcoder} models using the \emph{Mercury} benchmark. Each model was executed 50 times to ensure score robustness. The y-axis in the resulting histogram represents the frequency of observations within each bin.
        \vspace{-1em}
    }
    \label{fig:score_distribution}
\end{figure*}

\vspace{-0.5em}
\subsection{Dataset Metadata}
\vspace{-0.5em}
The Mercury dataset is hosted on Huggingface: \url{https://huggingface.co/datasets/Elfsong/Mercury}. The Croissant Metadata can be found at~\url{https://huggingface.co/api/datasets/Elfsong/Mercury/croissant}.

\vspace{-0.5em}
\subsection{Legal Compliance}
\vspace{-0.5em}
\label{sec:legal_compliance}
In this study, we have curated a comprehensive dataset by gathering publicly accessible task descriptions and archived solutions from LeetCode~(\url{https://leetcode.com/problemset/}). We have ensured that our collection process is strictly limited to tasks available in the free domain, intentionally excluding any content that falls under the paid services of the platform. We abide by Fair Use~\cite{fairuse} (Section 107): \textit{``the fair use of a copyrighted work, including such use by ... scholarship, or research, is not an infringement of copyright''}, where fair use is determined by \textit{``the purpose and character of the use, including whether such use is of a commercial nature or is for nonprofit educational purposes''}. With the \emph{Mercury} dataset, we emphasize its strictly non-commercial nature and underscore its purpose: to facilitate and advance academic research. The \emph{Mercury} dataset is released under Creative Commons Attribution Non Commercial 4.0~\cite{cc_by_nc_4}.

\end{document}